\documentclass[%
 aip,
 amsmath,amssymb,
 reprint,
]{revtex4-1}

\usepackage{graphicx}
\usepackage{dcolumn}
\usepackage{bm}

\usepackage{siunitx}

\newcommand{\kB}{k_\mathrm{B}}
\newcommand{\mbr}{\mathbf{r}}
\newcommand{\mbx}{\mathbf{x}}
\newcommand{\mbE}{\mathbf{E}}
\newcommand{\mbm}{\mathbf{m}}
\newcommand{\mbp}{\mathbf{p}}

\newcommand{\mbpi}{\boldsymbol{\pi}}
\newcommand{\mbalpha}{\boldsymbol{\alpha}}

\newcommand{\bol}[1]{\mbox{\boldmath $#1$}}
\newcommand{\be}{\begin{equation}}
\newcommand{\ee}{\end{equation}}

\def\v{\bol}
\def\bk{\bol{k}}
\def\vae{\varepsilon}
 
\begin{document}

\title[Quantum third dielectric virial coefficient]{Path-integral
  calculation of the third dielectric virial coefficient of noble gases}

\author{Giovanni Garberoglio}
\email{garberoglio@ectstar.eu}
 \affiliation{
   European Centre for Theoretical Studies in Nuclear Physics and
        Related Areas (FBK-ECT*), Trento, I-38123, Italy.}
\affiliation{Trento Institute for Fundamental Physics and Applications
  (TIFPA-INFN), Trento, I-38123, Italy.}

\author{Allan H. Harvey}
\email{allan.harvey@nist.gov}
\affiliation{Applied Chemicals and Materials Division, National Institute of
  Standards and Technology, Boulder, CO 80305, USA.}

\author{Bogumi\l{} Jeziorski}
\email{jeziorsk@chem.uw.edu.pl}
\affiliation{Faculty of Chemistry, University of Warsaw, Pasteura 1, 02-093
  Warsaw, Poland.}

\date{\today}

\begin{abstract}
We present the first framework for fully quantum calculation of the third
dielectric virial coefficient $C_\varepsilon(T)$ of noble gases, including
exchange effects.  The quantum effects are taken into account with the
path-integral Monte Carlo method.  Calculations employing state-of-the-art
pair and three-body potentials and pair polarizabilities yield results
generally consistent with the few scattered experimental data available for
helium, neon, and argon, but rigorous calculations with well-described
uncertainties will require the development of surfaces for the three-body
nonadditive polarizability and the three-body dipole moment.  
The framework developed here will enable new
approaches to primary temperature and pressure metrology based on
first-principles calculations of gas properties.
\end{abstract}

\maketitle

\section{Introduction}

Just as the deviation of a gas's thermodynamic behavior from that of an ideal gas is described by the familiar virial expansion, the dielectric virial expansion describes the low-density behavior of the static dielectric constant $\varepsilon$.
The Clausius--Mossotti function for a low-density gas of identical
molecules can be expanded in powers of molar density
$\rho$ as:
\begin{equation}
  \frac{\varepsilon - 1}{\varepsilon + 2} = \rho \left( A_\varepsilon + B_\varepsilon\rho
  + C_\varepsilon\rho^2 + \ldots \right)
  = \rho A_\varepsilon \left( 1 + b\rho + c\rho^2 + \ldots \right) \ ,
\label{eq:Vir:eps}
\end{equation}
where for non-polar molecules $A_\varepsilon$ is proportional to the trace of the static
polarizability of the isolated molecule.  Both the lower-case and
upper-case coefficients in Eq.~(\ref{eq:Vir:eps}) are sometimes called
``dielectric virial coefficients'' in the literature; in this work we use
the upper-case quantities where $B_\varepsilon$ is the second dielectric
virial coefficient, $C_\varepsilon$ is the third dielectric virial
coefficient, etc.

The dielectric virial expansion has seen increasing use in precision
metrology, particularly for fundamental measurements of pressure and of the
thermodynamic temperature.  For example, in dielectric-constant gas
thermometry,\cite{Gaiser_2015,Gaiser_2017,Gaiser_2017b,Gaiser_2020a}
capacitance measurements on noble gases (where the dielectric virial
coefficients are relatively small) are able to determine the thermodynamic
temperature with uncertainties smaller than 1~mK.  Another example is a
primary pressure standard up to 7 MPa based on measuring the static
dielectric constant of helium;\cite{Gaiser20} the uncertainty in
$C_\varepsilon$ was one of the two largest contributors to the uncertainty
budget of the standard.  A related expansion for the refractivity is also
used in refractive-index gas thermometry, and in most implementations the
frequency is close enough to the static limit that the appropriate
coefficients to use are those in the dielectric virial
expansion, with an additional term proportional to the magnetic
susceptibility.\cite{Piszczatowski15,Rourke_2019,Gao_2020,Ripa_2021,
  Rourke_2021}
When the static limit is not accurate enough -- e.g., for refractive-index gas
metrology at optical frequencies -- the frequency dependence of the
coefficients in Eq.~(\ref{eq:Vir:eps}) is also
required.~\cite{Piszczatowski15,Garberoglio20:Beps}

Particularly
in the case of helium where the polarizability of the isolated atom (and
therefore $A_\varepsilon$) can be calculated with extraordinary
accuracy,\cite{Puchalski20} much of the interest lies in first-principles
calculations of the virial coefficients in order to allow, for example,
calculation of pressure from a dielectric measurement without any need for
external calibration.

The second dielectric virial coefficient $B_\varepsilon$ depends only on
temperature for a given fluid; for monatomic species it can be calculated
from the interatomic potential-energy curve and the nonadditive two-body
polarizability.  Similarly, $C_\varepsilon$ requires knowledge of the
three-body potential, the three-body polarizability, and the three-body
dipole moment.  While the classical
calculation of $B_\varepsilon$, and to a lesser extent $C_\varepsilon$, is
fairly straightforward if the potentials, polarizabilities and the dipole
moment are known, helium is light enough that quantitative accuracy
requires the inclusion of quantum effects, and at low temperature quantum
effects can be significant for neon and to a lesser extent argon.

Recently, we reported fully quantum calculations of the second dielectric
and refractivity virial coefficients of helium, neon, and argon based on
state-of-the-art pair potentials and pair
polarizabilities.\cite{Garberoglio20:Beps} The calculations were performed
with the venerable wavefunction-based method for quantum two-body
problems,\cite{Hirschfelder:54} but the work also introduced a
path-integral approach for the quantum calculation of $B_\varepsilon$,
showing that the two independent approaches agreed (they also agreed with
semiclassical results\cite{Song_2020} at the high and moderate temperatures
where the semiclassical approach to quantum effects is valid).  The
advantage of the path-integral approach is that, just as in the case of the
thermodynamic third virial
coefficient,\cite{Garberoglio2009b,Garberoglio2011} it can be extended to
compute the third dielectric virial coefficient where no exact quantum
solution is known.

The extension of the path-integral method to the quantum calculation of
$C_\varepsilon$ is the main topic of this paper.  After a review of
previous calculations, we will present a derivation of the dielectric
virial expansion, including expressions for $B_\varepsilon$ and
$C_\varepsilon$.  We will then derive a path-integral formulation for
$C_\varepsilon$, which will include all quantum effects, including
exchange.  Calculations will then be performed for $^4$He, neon, and
argon, and comparisons will be made with the limited experimental data
available.

\section{Previous Calculations}
The only previous attempt to calculate $C_\varepsilon$ completely from first principles was in a 1974 paper by Heller and Gelbart,\cite{Heller74}
who reported a value of \SI{-0.716}~{cm${}^9$~mol${}^{-3}$} for helium at ``room temperature.''
This value has been used, either by itself or in combination with reported experimental values, in many metrology applications,\cite{Puchalski16,Rourke_2017,Gaiser_2019,Gao_2020,Gaiser20} including some that are not near room temperature.
However, there are several problems with general use of the value of Heller and Gelbart:

\begin{itemize}
\item 
It ignores temperature dependence of $C_\varepsilon$.  There is a large
temperature dependence both for helium's thermodynamic third virial
coefficient\cite{Garberoglio2011} and its second dielectric virial
coefficient.\cite{Garberoglio20:Beps} Assuming $C_\varepsilon$ to be
independent of temperature is clearly unjustified.
\item
It is an entirely classical calculation.  This should not introduce much
error for applications near room temperature, but quantum effects are
likely to be significant at cryogenic temperatures.
\item
The input to the calculation was primitive by today's standards.  For
simplicity, Heller and Gelbart assumed the intermolecular potential to be
that of a hard sphere.  The pair polarizability was taken from a 1973
paper\cite{OB_1973} that used a relatively low level of theory by modern
standards.  The three-body potential was assumed to be zero, and the
three-body polarizability was approximated in a simple way.
\end{itemize}

The 1980 paper of Alder \textit{et al.}\cite{Alder80} contains values of $C_\varepsilon$ calculated classically for helium and argon, but upon closer inspection it is evident that a parameter in the pair polarizability for each substance was arbitrarily adjusted to obtain agreement with some experimental data.
We note that the sources of experimental data they used\cite{Vidal76,Lallemand77} report values of $B_\varepsilon$ that are inconsistent with recent results for helium, neon, and argon.\cite{Garberoglio20:Beps}

These two calculational papers approximated the three-body polarizability with somewhat different versions of a ``superposition approximation,'' stated to be valid in the limit of large separations.
The form of this approximation will be discussed further in Section~\ref{sec:superpos}.

\section{The virial expansion of the dielectric constant}

In the literature, one can find several mutually incompatible    expressions
 for $C_{\vae}(T)$~\cite{Hill58,Mos1,GG2}.  In the following, we will present the correct one, 
highlighting similarities and points of departure from the others.

\subsection{Electric fields and polarization in homogeneous isotropic media}
\label{sec:Emedia}

We consider a region of volume $V$ in a quantum gas
of identical polarizable particles subject to an externally
applied electric field (generated, for example, by a distribution
$\varrho(\mbx)$ of electric charges on some external conductors).
Let us denote by $\v{E}_0$ the electric
field in the region $V$ coming from the sum of the electric field generated
by $\varrho(\mbx)$ and the polarization of the
gas {\em external} to the volume $V$. We assume that the volume
$V$ contains enough particles to justify a statistical
description (that is, temperature and chemical potential can be defined),
but small enough to neglect the spatial variation of $\v{E}_0$ within it.
We further assume that we deal with a linear
dielectric, that is the electric field is everywhere weak enough so that the
polarization density $\v{P}$ within $V$, defined as the dipole moment per
unit of volume, depends linearly on the applied field.
The dielectric constant $\vae$  (the relative electric permittivity)  for   
this system is defined by the relation~\cite{JDJ}
\be 
 \v{P}=\frac{\vae -1}{4\pi}\v{ E},
\label{eq:epsdef}
\ee where $\v{E}$ is the macroscopic electric field in the medium (the
Maxwell field).  This definition is equivalent to the statement that $\v{
  D}=\vae \v{E}$, where $\v{D}=\v{E} +4\pi\v{P}$ is the electric
displacement (electric induction).~\cite{JDJ}
In this case, $\v{ E}$  is the sum of the field $\v{ E}_0$ and the
average value of the field generated by the polarized medium within $V$. As
the dielectric constant is independent on the shape of the arbitrarily
chosen small volume $V$, we can assume that $V$ is spherical. In this case,
the latter field is equal to $ - (4\pi/3) \v{P} $,~\cite{JDJ} and we have
\be
\v{\  E}= \v{ E}_0-\frac{4\pi}{3}\v{P}.
\label{eq:avgEbar}
\ee
Combining Eq.~(\ref{eq:epsdef}) with Eq.~(\ref{eq:avgEbar}) leads to 
\be
\v{P} =  \frac{3}{4\pi}\, \frac{\vae-1}{\vae +2}\, \v{E}_0.
\label{eq:PE0}
\ee  
At the end of the calculation, we can take the
thermodynamic limit $V\rightarrow \infty$, and our system becomes
equivalent to an infinite homogeneous gas in an external field $\v{E}_0$,
i.e., we neglect any boundary effects.
For an atomic gas, the vectors $\v{E}_0$, $\v{P}$, $\v{E}$, and
$\v{D }$ are parallel so that $\v{E}_0$=$E_0 \bk$, $\v{P}$=$P \bk$,
$\v{E}$=$E \bk$, and $\v{D}$=$D \bk$, where $\bk$ is the unit vector
parallel to the external field $\v{E}_0$.

\subsection{Quantum statistical mechanics of the dielectric response}

In view of Eq.~(\ref{eq:PE0}) and Eq.~(\ref{eq:Vir:eps}), the dielectric
viral coefficients can be obtained by expanding $4\pi P/(3E_0)$ in powers
of particle number density $\rho$.  Application of classical~\cite{Hill58} or
quantum~\cite{Ceps_forth} statistical mechanics results in the following expansion
for $P$
\begin{equation}
  P = \alpha_1 \rho E_0 - \kB T \sum_{n=2}^\infty \left(\frac{1}{n-1}\right)
  \left. \frac{\partial B_n(T,E_0)}{\partial E_0} \right|_{E_0=0}\rho^n,
  \label{eq:PHill}
\end{equation}
where $\alpha_1$ is the atomic polarizability, $ T$ is the temperature, and
$B_n(T,E_0)$ is the usual $n$th density virial coefficient for an infinite
system in the external static and uniform electric field of magnitude $E_0$.

It should be noted that the formula for $P$ given by Eq. (24) in
Ref.~\onlinecite{Hill58} is formally identical with Eq.~(\ref{eq:PHill})
except that the external field strength $E_0$ is replaced by the electric
displacement $D$, which leads to incorrect expressions for the second and
higher dielectric virial coefficients.

The density virials $B_n(T,E_0)$ are given as the $V \to \infty$
combinations of the functions $Z_N(V,T,E_0)$ defined as
\begin{equation}
  \frac{Z_N(T,E_0)}{N!} = \frac{Q_N(V,T,E_0) V^N}{Q_1(V,T,E_0)^N},
  \label{eq:ZN}
\end{equation}
where $Q_N(V,T,E_0)$ are the canonical partition functions of $N$ particles
in volume $V $ in the presence of an external uniform electric field of
magnitude $E_0$. In classical mechanics, $Q_N(V,T,E_0)$ is the phase
integral over the Boltzmann factor $\mathrm{e}^{-\beta H(V,N,E_0)}$ divided by $N!
h^{3N}$, where $\beta= 1/\kB T$, $h$ is the Planck constant, and $H(V,N,E_0)$
is the classical Hamiltonian of the system.  In quantum mechanics,
$Q_N(V,T,E_0)$ is the trace of the Boltzmann operator $\mathrm{e}^{-\beta
  H(V,N,E_0)}$ in the bosonic or fermionic Hilbert space, where now
$H(V,N,E_0)$ stands for the quantum Hamiltonian.
In particular, one has~\cite{Hirschfelder:54}
\begin{eqnarray}
  B_2 &=& -\frac{1}{2V} \left( Z_2 - V^2 \right) \\
  B_3 &=& \frac{\left( Z_2 - V^2 \right)^2}{V^2} - \frac{1}{3V} \left(
  Z_3 - 3 Z_2 V + 2 V^3 
  \right).
\end{eqnarray}
Since the derivation of Eq.~(\ref{eq:PHill}) has been performed in the grand
canonical ensemble, we think of the volume $V$ as a part of the volume of an
experimental apparatus. We assume that $V$ is large enough to contain
enough atoms to fulfill the requirements of the thermodynamic limit. Notice
that the quantities $B_n(T,E_0)$ are finite in the usual \mbox{$V\rightarrow
\infty$} limit.

We now follow Moszynski {\em et al.}~\cite{Mos1} and substitute in
Eq.~(\ref{eq:PHill}) the equivalence
\begin{equation}
  \frac{\partial B_n(T,E_0)}{\partial E_0} = \frac{\partial^2
    B_n(T,E_0)}{\partial^2 E_0} E_0,
\end{equation}
which is valid in the $E_0 \to 0$ limit since $B_n(T,E_0)$ depends
quadratically on $E_0$,\cite{Hill58} thus obtaining
\begin{equation}
  \frac{P}{E_0} = \alpha_1 \rho - \kB T \sum_{n=2}^\infty \left(\frac{1}{n-1}\right)
  \frac{\partial^2 B_n(T,E_0)}{\partial^2 E_0} \rho^n.
\end{equation}
Recalling Eq.~(\ref{eq:PE0}), we finally arrive at
\begin{eqnarray}
  \frac{\varepsilon-1}{\varepsilon+2} &=& \frac{4\pi}{3} \left[
  \alpha_1 \rho - \kB T \sum_{n=2}^\infty \left(\frac{1}{n-1}\right)
  \frac{\partial^2 B_n(T,E_0)}{\partial^2 E_0} \rho^n
  \right] \label{eq:dielectric_expansion} \\
  &\equiv& A_\varepsilon \rho + B_\varepsilon \rho^2 + C_\varepsilon \rho^3
  + \ldots \label{eq:CoefDef} \\ 
  A_\varepsilon &=& \frac{4 \pi \alpha_1}{3} \\
  B_\varepsilon &=& \frac{2 \pi \kB T}{3 V}
  \frac{\partial^2 Z_2(V,T,E_0)}{\partial E_0^2} \label{eq:Beps} \\
  C_\varepsilon &=& -\frac{2 \pi \kB T}{3} \left[
    \frac{2}{V^2} \left( \frac{\partial Z_2}{\partial E_0} \right)^2 +
    \frac{2(Z_2-V^2)}{V^2} \frac{\partial^2 Z_2}{\partial E_0^2}
    \right. + \nonumber \\
  & & \left. -\frac{1}{3V} \left(
  \frac{\partial^2 Z_3}{\partial E_0^2} - 3 V \frac{\partial^2
    Z_2}{\partial E_0^2}
  \right) \right] \label{eq:Ceps}
\end{eqnarray}
which is the virial expansion that we will use in this paper, with the
coefficients defined in Eq.~(\ref{eq:CoefDef}).
Keeping the lowest term in $\rho$ recovers the Clausius--Mossotti equation,
whereas the coefficient of the term $\rho^2$ that we obtain is in perfect
agreement with the results reported in Refs.~\onlinecite{Mos1} and
\onlinecite{GG2}.

As already mentioned our expression for $P$, and consequently
for the dielectric virial coefficients differs from that
of Hill~\cite{Hill58}.  In fact, his expression for $P$ is incorrect
even for the ideal gas as it does not lead to the Clausius--
Mossotti equation in this case, see Eqs. (32)-(34) in
Ref.~\onlinecite{Hill58}.
In a later paper,~\cite{Hill59}  Hill  proposed to correct
his  expression for $P$ by substituting the Maxwell field
$E$ for the electric displacement $D$. The resulting expression
for $P$ remains incorrect.

Although we obtain the same formula for $B_\varepsilon$ as
Moszynski {\em et al.},~\cite{Mos1} our expression for $C_\varepsilon$
differs from theirs, because they mistakenly use
Eq.~(\ref{eq:epsdef}) with $E = E_0$ in all the terms of the
right-hand side of Eq.~(\ref{eq:PHill}) except the first, for which
their considerations are equivalent to ours.
Finally, we notice that our expression for the third dielectric virial
coefficient is equivalent, in the classical limit, to the
one derived in Ref.~\onlinecite{GG2}.
Equation (\ref{eq:Ceps}) can also be obtained starting from the
fundamental equations of quantum statistical mechanics, and we will present
this alternative derivation in a forthcoming work.~\cite{Ceps_forth}

\subsection{Hamiltonians in an external field}
\label{sec:partfun}

In the case of polarizable atoms, the partition functions $Q_N(V,T,E_0)$
appearing in Eq.~(\ref{eq:ZN}) are obtained using an $N$-particle
Hamiltonian function $H(N) = H_0(N) + \Delta H_1(N) + \Delta H_2(N)$, which
is the sum of the 
Hamiltonian $H_0(N)$ of $N$ atoms without the external field plus
two contributions $\Delta H_1(N)$ and $\Delta H_2(N)$ describing the
linear and quadratic interaction with the external field,
respectively (terms with higher order in $\mbE_0$ do not contribute to the
dielectric constant). These three functions are given by 
\begin{eqnarray}
  H_0(N) &=& \sum_{i=1}^N \frac{\mbpi_i^2}{2m} +
  \sum_{i<j} u_2(i,j) + \sum_{i<j<k} u_3(i,j,k) + \ldots,
  \label{eq:H0} \\
  \Delta H_1(N) &=& - \left(
  \sum_{i=1}^N \mbm_1(i) + \sum_{i<j} \mbm_2(i,j) + \right . \nonumber \\
  & & \left.
  \sum_{i<j<k} \mbm_3(i,j,k) + \ldots  \right) \cdot \mbE_0 \label{eq:DH1} \\
  \Delta H_2(N) &=& -\frac{1}{2} \mbE_0 \cdot \left(
  \sum_{i=1}^N \mbalpha_1(i) + \sum_{i<j} \mbalpha_2(i,j) + \right.
  \nonumber \\
  & & \left. 
  \sum_{i<j<k} \mbalpha_3(i,j,k) + \ldots \right) \cdot \mbE_0,
  \label{eq:DH2}
\end{eqnarray}
where $m$ is the mass of the atoms, $\mbpi_i$ is the momentum of the $i$-th
atom, $u_k(1,\ldots,k)$ is the irreducible $k$-body potential,
$\mbm_k(1,\ldots,k)$ is the dipole moment and $\mbalpha_k(1,\ldots,k)$ is
the $k$-body induced polarizability, which is a $3 \times 3$ matrix.  For
noble gases, $\mbm_1(i)$ and $\mbm_2(i,j)$ are identically zero, but a
configuration of three atoms can have a permanent dipole, hence
$\mbm_3(i,j,k)$ is in general not zero.~\cite{Martin74,Bruch78}

\subsection{The induced-dipole--induced-dipole model for the polarizability}
\label{sec:superpos}

In general, the quantities $\mbm_k(1,\ldots,k)$ and
$\mbalpha_k(1,\ldots,k)$ appearing in Eqs.~(\ref{eq:DH1}) and
(\ref{eq:DH2}) must be obtained by {\em ab initio} calculations of the
ground-state energy of a cluster of $n$ atoms in the presence of an
external field. Presently, only $\mbalpha_2$ is known with sufficient
precision to enable highly accurate calculations of
$B_\varepsilon$~\cite{Garberoglio20:Beps} for helium, neon, and argon.  In
the case of $\mbm_3$, a first-principles parameterization for noble gases,
valid in the limit of large distances, has been developed by Li and
Hunt.~\cite{Li97} Its accuracy is unknown, but as will be discussed in
Sec.~\ref{results}, the contribution of $\mbm_3$ to $C_\varepsilon(T)$
appears to be relatively small.

To the best of our knowledge, no first-principles surface for $\mbalpha_3$
has been published in the literature. As we will discuss below, the
contribution from the three-body polarizability to $C_\varepsilon(T)$ turns
out to be substantial and for this reason it is worthwhile to briefly
investigate models for the polarizability that lead to approximate
expressions for $\mbalpha_3$ as a function of the lower-order
polarizabilities. One of the most used models assumes that the dipole
moment $\mbp_i$ of atom $i$ depends on the the total field acting on it
\begin{equation}
  \mbp_i = \alpha_1 \mbE_\mathrm{tot}(\mbr_i),
\end{equation}
and that the total field is given by the sum of the externally applied
field $\mbE_0$ and the contribution of the fields generated by the induced
dipoles
\begin{equation}
  \mbE_\mathrm{tot}(\mbr_i) = \mbE_0 + \sum_{j \neq i}
  \mathsf{T}(\mbr_i - \mbr_j) \mbp_j(\mbr_j),
\end{equation}
where the tensor $\mathsf{T}(\mbr)$ is given by
\begin{equation}
  \mathsf{T}_{\alpha \beta}(\mbr) = \frac{3 r_\alpha r_\beta}{r^5} - \frac{\delta_{\alpha\beta}}{r^3},
\end{equation}
where $r_\alpha$ is the $\alpha$-th component of the vector $\mbr$.

The solution of these equations in the case of three particles~\cite{Buck91}
shows that one has
\begin{eqnarray}
  \mbalpha_2(\mbr_1,\mbr_2) &\simeq& 2 \alpha_1^2 \mathsf{T}(\mbr_{12}) \\
  \mbalpha_3(\mbr_1,\mbr_2, \mbr_3) &\simeq& 2 \alpha_1^3 \left[
  \mathsf{T}(\mbr_{12}) \mathsf{T}(\mbr_{23}) + \right. \nonumber \\
  & & \mathsf{T}(\mbr_{23}) \mathsf{T}(\mbr_{31}) + \nonumber \\
  & & \left. \mathsf{T}(\mbr_{31}) \mathsf{T}(\mbr_{12})
  \right],
\end{eqnarray}
where we have defined $\mbr_{ij} = \mbr_i - \mbr_j$. The last result can
be generalized by the so-called {\em   superposition
  approximation}~\cite{Gelbart74,Heller74,Alder80} for 
$\mbalpha_3$ as a function of $\mbalpha_2$,
\begin{widetext}
\begin{equation}
  \mbalpha_3(\mbr_1,\mbr_2,\mbr_3) = \frac{1}{2 \alpha_1} \left[
  \mbalpha_2(1,2) \mbalpha_2(2,3) +
  \mbalpha_2(2,3) \mbalpha_2(3,1) +
  \mbalpha_2(3,1) \mbalpha_2(1,2) \right],
  \label{eq:superposition}
\end{equation}  
\end{widetext}
where we denoted $\mbalpha_2(i,j) = \mbalpha_2(\mbr_i, \mbr_j)$.
Similar approximations appeared in the literature: Heller and
Gelbart~\cite{Heller74} approximated the $zz$ component of $\mbalpha_3$
similarly to Eq.~(\ref{eq:superposition}), but considered it as given by the
product of the $zz$ components of the two $\mbalpha_2$ tensors, and did not
consider the factor of $1/2$. Our definition of the superposition
approximation is the same as that used by Alder \textit{et al.}\cite{Alder80}
Equation~(\ref{eq:superposition}) agrees asymptotically at large
interatomic distances with the exact asymptotics of the three-body
polarizability derived by Champagne et al.~\cite{Champagne2k}

\subsection{Structure of the pair and three-body polarizabilities}

In general, the pair polarizability $\mbalpha_2(\mbr)$ is written as
\begin{equation}
  \mbalpha_2(\mbr)_{\alpha\beta} = \alpha_\mathrm{iso}(r) \delta_{\alpha\beta} +
  \frac{\alpha_\mathrm{aniso}(r)}{3} \mathsf{t}_{\alpha\beta}(\mbr)
  \label{eq:alpha_2}
\end{equation}
where $\mathsf{t}_{\alpha\beta}(\mbr) = r^3
\mathsf{T}_{\alpha\beta}(\mbr)$. The quantity $\alpha_\mathrm{iso}(r)$ and 
$\alpha_\mathrm{aniso}(r)$ are known as the average trace of the pair-induced
polarizability and the anisotropic component, respectively.
As we will see below, the relevant quantities involved in the calculation
of the dielectric virial coefficients are the traces of $\mbalpha_n$
defined in Eq.~(\ref{eq:DH2}). A straightforward calculation using
Eqs.~(\ref{eq:alpha_2}) and (\ref{eq:superposition}) shows that
\begin{widetext}
  \begin{eqnarray}
    \frac{1}{3}\mathrm{tr}\left[\mbalpha_3(r_{12}, r_{13}, r_{23})\right] &=&
    \frac{1}{2 \alpha_1} \left[
      \alpha_\mathrm{iso}(r_{12}) \alpha_\mathrm{iso}(r_{23}) +
     \frac{\alpha_\mathrm{aniso}(r_{12}) \alpha_\mathrm{aniso}(r_{23})}{3} \left( \cos^2\theta_2 - \frac{1}{3} \right)
      + \right.  \nonumber \\
     & &       \alpha_\mathrm{iso}(r_{13}) \alpha_\mathrm{iso}(r_{32}) +
      \frac{\alpha_\mathrm{aniso}(r_{13}) \alpha_\mathrm{aniso}(r_{32})}{3} \left( \cos^2\theta_3 - \frac{1}{3} \right)
      + \nonumber \\
& &  \left.     \alpha_\mathrm{iso}(r_{12}) \alpha_\mathrm{iso}(r_{13}) +
      \frac{\alpha_\mathrm{aniso}(r_{12}) \alpha_\mathrm{aniso}(r_{13})}{3} \left( \cos^2\theta_1 - \frac{1}{3} \right)
      \right],
\label{eq:super_a3}    
  \end{eqnarray}
\end{widetext}
where $\theta_i$ is the angle at particle $i$ in the triangle having as
vertices the three particles considered.

\section{The path-integral formulation}

In general, the partition functions involved in the definition of $Z_N$
(see Eq.~(\ref{eq:ZN})) can be written as
\begin{equation}
  Q_N(V,T,E_0) = \frac{1}{N!} \sum_{i,\sigma} \langle i | \mathrm{e}^{-\beta H(N)}
  P_\sigma |i \rangle,
  \label{eq:QN}
\end{equation}
where the states $|i\rangle$ denote a complete basis set in the Hilbert
space of $N$ atoms, $\sigma$ are permutations of $N$ objects, and
$P_\sigma$ is the operator representing the permutation in the Hilbert
space, weighted with the sign of the permutation in the case of
fermions. At high temperatures, when the de~Broglie
thermal wavelength $\Lambda = h / \sqrt{2 m \kB T}$ is much smaller than
the hard-core radius of the atoms, only the term where $\sigma$ is the
identity permutations contributes to Eq.~(\ref{eq:QN}). In this case, the
bosonic or fermionic nature of the quantum particles is not apparent and
particles behave as distinguishable (Boltzmann limit); quantum effects
appear only as a consequence of Heisenberg uncertainty (diffraction
limit). At low temperatures, where the thermal wavefunctions begin to
overlap significantly ($T \leq 4$~K for ${}^4$He), the terms with
nontrivial permutations in Eq.~(\ref{eq:QN}) become appreciable and the
exchange effects (related to the bosonic or fermionic nature of the
particles under consideration) become significant.

\subsection{Boltzmann contribution}

We will use the path-integral formulation of quantum statistical
mechanics~\cite{FH} to obtain an expression of $C_\varepsilon(T)$ useful in actual
calculations. From Eq.~(\ref{eq:Ceps}), we need to consider the derivatives
of both $Z_2$ and $Z_3$ with respect to an external field.
Using Eq.~(\ref{eq:QN}),~\cite{Garberoglio20:Beps,Garberoglio2011a,Garberoglio2011aerr}
$Z_2$ can be written as
\begin{eqnarray}
  Z_2 &\equiv& Z_2^: + Z_2^| \label{eq:Z2}\\
  Z_2^: &=& \int
  \langle \mbr_1^{(1)} \mbr_2^{(1)} | \mathrm{e}^{-\beta H(2)} | \mbr_1^{(1)} \mbr_2^{(1)} \rangle
  \mathrm{d}\mbr_1^{(1)} \mathrm{d}\mbr_2^{(1)} \\
  Z_2^| &=& \frac{(-1)^{2I}}{2I+1}
  \int \langle \mbr_1^{(1)} \mbr_2^{(1)} | \mathrm{e}^{-\beta H(2)} | \mbr_2^{(1)}
  \mbr_1^{(1)} \rangle
    \mathrm{d}\mbr_1^{(1)} \mathrm{d}\mbr_2^{(1)}
\end{eqnarray}
where $\mbr_1^{(1)}$ and $\mbr_2^{(1)}$ are the coordinates of the atoms
and $I$ is their nuclear spin ($I=0$ for ${}^4$He, ${}^{20}$Ne, and
${}^{40}$Ar, $I=1/2$ for ${}^3$He). The first term in Eq.~(\ref{eq:Z2}) ($Z_2^:$) is known as the Boltzmann
term, whereas the second ($Z_2^|$) is called the exchange term. The
superscript visually represents the kind of permutation 
$\sigma$ that is involved in the definition of the partition function.
The superscript $(1)$ in the positions has been introduced for convenience in the
path-integral formulation of $Z_2$, which is based on the Trotter expansion
\begin{equation}
  \mathrm{e}^{-\beta H(2)} \simeq
  \left( \mathrm{e}^{-\beta T(2)/P} \mathrm{e}^{-\beta V(2)/P} \mathrm{e}^{-\beta
    \Delta H(2)/P} \right)^P,
  \label{eq:Trotter}
\end{equation}
for sufficiently large $P$, where $T(2)$ is the kinetic energy of the two
atoms, $V(2)$ their potential energy ($T(2) + V(2) = H_0(2)$, see
Eq.~(\ref{eq:H0})) and $\Delta H_2(2)$ the interaction energy with the
external field $E_0$, from Eq.~(\ref{eq:DH2}). We will assume, without
losing generality, that $E_0$ is
directed along the $z$ axis. Substituting
Eq.~(\ref{eq:Trotter}) into Eq.~(\ref{eq:Z2}) and inserting $P-1$
completeness relations, one obtains the expression
\begin{eqnarray}
  \frac{Z_2^:}{V} &=& \int \left\langle \exp\left(-\beta
  \overline{V_2^:}(\mbr) +
  \frac{\beta}{2} \overline{\alpha_{2}^:}_{,zz}(\mbr) E_0^2 \right) \right\rangle ~
  \mathrm{d}\mbr \label{eq:Z2B}\\
  \overline{V_2^:}(\mbr) &=& \frac{1}{P} \sum_{i=1}^P u_2(\mbr^{(i)}) \\
  \overline{\alpha_{2}^:}_{,zz}(\mbr) &=& \frac{1}{P} \sum_{i=1}^P \alpha_{2,zz}(\mbr^{(i)}),  
\end{eqnarray}
where $\mbr^{(i)} = \mbr_2^{(i)} - \mbr_1^{(i)}$ and we have denoted $\mbr
= \mbr_2^{(1)} - \mbr_1^{(1)}$. The average $\langle \cdots \rangle$ in
Eq.~(\ref{eq:Z2B}) is performed over two distribution functions
$\Pi_k(\Delta \mbr_k^{(i)})$ ($k=1,2$)~\cite{Garberoglio2008} that depend
on the $P$ quantities $\Delta\mbr_k^{(i)} = \mbr_k^{(i+1)} - \mbr_k^{(i)}$,
with the understanding that $\mbr_k^{(P+1)} = \mbr_k^{(1)}$ (notice that
this condition implies that $\Delta\mbr^{(P)}$ is opposite to the sum of
all the other $\Delta\mbr^{(i)}$).  The distribution functions $\Pi_k$ are
given by~\cite{Garberoglio2008}
\begin{equation}
  \Pi(\Delta\mbr^{(i)}; P) =
  \Lambda^3 \left(\frac{P^{3/2}}{\Lambda^3}\right)^P
  \exp\left(-\frac{\pi P}{\Lambda^2} \sum_{i=1}^P
  \left| \Delta\mbr^{(i)}\right|^2
  \right),
  \label{eq:Pi}
\end{equation}
which can be interpreted as the probability distribution of a
classical closed ring polymer with $P$
monomers.~\cite{FH,Garberoglio2008,Garberoglio20:Beps}

The first derivative with respect to $E_0$ of Eq.~(\ref{eq:Z2B}) produces
\begin{equation}
  \frac{1}{V} \frac{\partial Z_2^:}{\partial E_0} = \int
  \beta E_0 \left\langle \overline{\alpha_{2}^:}_{,zz}(\mbr)
  \exp\left({-\beta \overline{V_2^:}(\mbr) +
    \frac{\beta}{2} \overline{\alpha_{2}^:}_{,zz}(\mbr) E_0^2}\right)  
  \right\rangle,
\end{equation}
so that the first term in Eq.~(\ref{eq:Ceps}) is seen to be zero when
evaluated at zero external field. The second derivative at zero field
becomes
\begin{equation}
  \frac{1}{V} \left. \frac{\partial^2 Z_2^:}{\partial E_0^2}\right|_{E_0=0}
  = \beta \int
  \left\langle \overline{\alpha_{2}^:}_{,zz}(\mbr)
  \mathrm{e}^{-\beta \overline{V_2^:}(\mbr)}
  \right\rangle ~ \mathrm{d}\mbr,
  \label{eq:dZ2B}
\end{equation}
and hence we get directly the Boltzmann contribution to the second
dielectric virial coefficient, that is
\begin{equation}
  B_\varepsilon^:(T) = \frac{8 \pi^2}{3} \int
  \left\langle \overline{\alpha_\mathrm{iso}^:}(r)
  \mathrm{e}^{-\beta \overline{V_2^:}(r)}
  \right\rangle
  ~ r^2 \mathrm{d}r,
\end{equation}
accounting for the rotational invariance, that is $\overline{\alpha_{2}^:}_{,zz} =
\left( \overline{\alpha_{2}^:}_{,xx} + \overline{\alpha_{2}^:}_{,yy} +
\overline{\alpha_{2}^:}_{,zz} \right)/3 = 
\overline{\alpha_\mathrm{iso}^:}$. This is the same equation as derived
in Ref.~\onlinecite{Garberoglio20:Beps}.

The same considerations apply to the calculation of the second derivative
of $Z_3$, which is obtained as the sum of three contributions, due to the
three possible permutations of three objects: the identity which results in
the Boltzmann component (which will be denoted by the symbol $\therefore$);
the permutations of any two particles (which are three in total and will be
denoted by $\cdot |$); and the cyclic permutations (two, denoted by
$\triangle$).~\cite{Garberoglio2008} 
For the Boltzmann part, one gets
\begin{equation}
  \frac{1}{V} \left. \frac{\partial^2 Z_3^\therefore}{\partial E_0^2} \right|_{E_0=0}
  = \beta \int \left\langle
  \left(\frac{\beta \left| \overline{\mbm_3^\therefore} \right|^2}{3} +
  \overline{A_3^\therefore} \right) \mathrm{e}^{-\beta
    \overline{V_3^\therefore}} \right\rangle ~ \mathrm{d}\mbr_1 \mathrm{d}\mbr_2
  \label{eq:dZ3}
\end{equation}
where
\begin{widetext}
  \begin{eqnarray}
    \overline{\mbm_3^\therefore}(\mbr_1, \mbr_2, \mbr_3) &=&
    \frac{1}{P} \sum_{p=1}^P
    \mbm_3\left(\mbr_1^{(p)}, \mbr_2^{(p)}, \mbr_3^{(p)}\right) \\
  \overline{A_3^\therefore}(\mbr_1, \mbr_2, \mbr_3) &=& \frac{1}{3P}\sum_{p=1}^P
  \mathrm{tr}\left(
  \alpha_3(\mbr_1^{(p)}, \mbr_2^{(p)}, \mbr_3^{(p)}) 
  + \sum_{i<j=1}^3 \alpha_2(\mbr_i^{(p)} - \mbr_j^{(p)})
  \right)  \\
  \overline{V_3^\therefore}(\mbr_1, \mbr_2, \mbr_3) &=&
  \frac{1}{P}\sum_{p=1}^P \left(
  u_3(\mbr_1^{(p)}, \mbr_2^{(p)}, \mbr_3^{(p)}) + 
  \sum_{i<j=1}^3 u_2(\mbr_i^{(p)} - \mbr_j^{(p)})  \right),
\end{eqnarray}  
\end{widetext}
and the average $\langle \cdots \rangle$ in Eq.~(\ref{eq:dZ3}) is performed
on three distribution functions $\Pi_k$ ($k=1,2,3$) analogous to what has
been done for $Z_2^:$.
Although these last equations have
been written using the coordinates of three particles, translational
invariance implies that one of the coordinates ($\mbr_1^{(1)}$, say) can be
placed at the origin, resulting in a factor of $V$ from the
integration. For this reason, the integration of Eq.~(\ref{eq:dZ3}) is
performed over the coordinates of the other two particles.
The final path-integral expression for the Boltzmann part of $C_\varepsilon$ is
\begin{widetext}
  \begin{equation}
    C_\varepsilon^\therefore(T) = \frac{2\pi}{3} \int 
    \left[ \frac{1}{3}\left\langle
      \left(\frac{\beta |\overline{\mbm_3^\therefore}|^2}{3} +
      \overline{A_3^\therefore} \right) \mathrm{e}^{-\beta
        \overline{V_3^\therefore}} - \sum_{i<j}
      \overline{\alpha_\mathrm{iso}^:}(\mbr_{ij}) \mathrm{e}^{-\beta
        \overline{V_2^:}(\mbr_{ij})} \right\rangle - 2 \langle \mathrm{e}^{-\beta
        \overline{V_2^:}(\mbr_{21})}-1 \rangle \langle \overline{\alpha_\mathrm{iso}^:}(\mbr_{31}) \mathrm{e}^{-\beta \overline{V_2^:}(\mbr_{31})}
      \rangle \right] ~ \mathrm{d}\mbr_2 \mathrm{d}\mbr_3,
  \label{eq:CepsB}
  \end{equation}
\end{widetext}
where $\mbr_{ij} = \mbr_i - \mbr_j$. The first average is taken over
three independent ring-polymer distributions, whereas the last two
averages are each taken over two independent ring-polymer distributions.
In the classical limit, the ring polymers shrink to a point and
Eq.~(\ref{eq:CepsB}) becomes the classical expression derived in
Ref.~\onlinecite{GG2}.

\subsection{Exchange effects}

Using an approach very similar to what has been outlined in the previous
section, one can derive path-integral expressions for the exchange
contributions. Many details can be found in our previous works (e.g.,
Refs.~\onlinecite{Garberoglio2011,Garberoglio2011err,Garberoglio2011a,Garberoglio2011aerr}),
so we just recall that the main effect of the permutation operators
$P_\sigma$ is to ``coalesce'' the $P$-bead ring polymers of the particles
involved in the permutation -- let us denote them by $n$ -- into a bigger
polymer with $nP$ beads, and at the same time introduce a multiplication
factor proportional to $\Lambda^{3(n-1)}$.

The ``coalescence'' of the ring polymers takes into account quantum
statistical effects due to the indistinguishability of the
particles. Qualitatively speaking, coalesced configurations will have a
sizable probability of being sampled as soon as the size of the ring
polymers -- which is, in turn, proportional to the de~Broglie thermal
wavelength $\Lambda$ (Ref.~\onlinecite{Garberoglio2012}) -- exceeds the size
of the repulsive core of the interatomic potential (which is usually between $2.5$
 and $3.5$~\AA), a condition that requires low temperatures. At higher
 temperatures, the dielectric virial coefficients are due entirely to the
 Boltzmann contribution, which takes into account the quantum nature of the
 particles only via the Heisenberg uncertainty (quantum diffraction effects).

After taking into account the effect of the permutation operator, one obtains 
\begin{eqnarray}
  Z_2^| &=& \frac{(-1)^{2I}}{2I+1} \frac{\Lambda^3}{2^{3/2}}
  \left\langle
  \exp\left(-\beta \overline{V^|} +
  \frac{\beta}{2} E_0 \overline{\alpha_2^|}_{,zz} E_0 \right)
  \right\rangle
  \label{eq:Z2|}
  \\
  \overline{V^|} &=& \frac{1}{2P} \sum_{i=1}^{2P} V(\mbx^{(i)}) \\
  \overline{\alpha_2^|}_{,zz} &=& \frac{1}{2P} \sum_{i=1}^{2P} \alpha_{2,zz}(\mbx^{(i)}),
\end{eqnarray}
where the coordinates $\mbx^{(i)}$ are defined so that $\mbx^{(i)} =
\mbr_1^{(i)}$ and $\mbx^{(i+P)} = \mbr_2^{(i)}$ for $i=1,\ldots,P$ . The
average
$\langle \cdots \rangle$
in Eq.~(\ref{eq:Z2|}) is performed over a distribution
$\Pi(\Delta \mbx^{(i)};2P)$ which is a function of the $2P$ coordinates
$\Delta\mbx^{(i)} = \mbx^{(i+1)}-\mbx^{(i)}$, with again the understanding
that $\mbx_k^{(2P+1)} = \mbx_k^{(1)}$.
Performing the derivatives, we obtain again that the first derivative at
zero field vanishes, whereas the second derivative can be written as
\begin{equation}
  \frac{\partial^2 Z_2^|}{\partial E_0^2} = \beta
  \frac{(-1)^{2I}}{2I+1} \frac{\Lambda^3}{2^{3/2}}
  \left\langle
  \overline{\alpha_2^|}_{,zz}
  \mathrm{e}^{-\beta \overline{V^|}}
  \right\rangle,
  \label{eq:dZ2|}
\end{equation}
which leads directly to the exchange term of $B_\varepsilon(T)$ discussed
in Ref.~\onlinecite{Garberoglio20:Beps}.

In the case of the third dielectric virial coefficient, there are several
contributions to exchange effects. The first comes from the terms involving
$Z_2$, when we express $Z_2 = Z_2^: + Z_2^|$, whereas other
contributions come from the term involving $Z_3$ that can be written as
\begin{equation}
  Z_3 = Z_3^\therefore + Z_3^{\cdot |} + Z_3^\triangle.
  \label{eq:Z3}
\end{equation}
The term $Z_3^{\cdot |}$ describes permutations of a single pair
(which are odd), whereas the term $Z_3^\triangle$ describes cyclic
permutations, which are even.
From Eqs.~(\ref{eq:Z2B}), (\ref{eq:dZ2B}), (\ref{eq:Z2|}),
(\ref{eq:dZ2|}), and (\ref{eq:Z3}), one obtains
\begin{widetext}
  \begin{eqnarray}
    C^{\cdot |}_\varepsilon(T) &=& \frac{(-1)^{2I}}{2I+1} \frac{2\pi}{3}
    \frac{\Lambda^3}{2^{3/2}}
    \int \mathrm{d}\mbr ~ \left[
    \left\langle \left(
    \frac{\beta \left|\overline{\mbm_3^{\cdot |}}\right|^2}{3} + 
    \overline{A_3^{\cdot |}} \right)
    \mathrm{e}^{-\beta \overline{V_3^{\cdot |}}}
    - \overline{\alpha_\mathrm{iso}^|} \mathrm{e}^{-\beta \overline{V_2^|}}
    \right\rangle
    - \nonumber \right. \\
    & &  \left.
    2 \langle \mathrm{e}^{-\beta \overline{V_2^|}} \rangle
    \langle \overline{\alpha_\mathrm{iso}^:} \mathrm{e}^{-\beta
      \overline{V_2^:}} \rangle 
    -
    2 \langle \overline{\alpha_\mathrm{iso}^|} \mathrm{e}^{-\beta \overline{V_2^|}} \rangle
    \langle \mathrm{e}^{-\beta \overline{V_2^:}} -1 \rangle \right]
    \\
    C^\triangle_\varepsilon(T) &=& \frac{2 \pi}{3}
    \frac{\Lambda^6}{(2I+1)^2} \left[
    \frac{2}{3^{5/2}} \left\langle
    \left(
    \frac{\beta \left| \overline{\mbm_3^\triangle} \right|^2}{3} + 
    \overline{A_3^\triangle} \right)
    \mathrm{e}^{-\beta \overline{V_3^\triangle}}
    \right\rangle
    -
    \frac{1}{4} \left\langle \mathrm{e}^{-\beta \overline{V_2^|}} \right\rangle
    \left\langle \overline{\alpha_\mathrm{iso}^|} \mathrm{e}^{-\beta \overline{V_2^|}} \right\rangle
    \right]
  \end{eqnarray}
\end{widetext}

\subsection{Details of the calculations}

In the following, we will present results for $C_\varepsilon(T)$ neglecting
exchange effects, so using Eq.~(\ref{eq:CepsB}) only. The main
reason is that fully \emph{ab initio} expressions for $\alpha_3$ and $\mbm_3$ are not known,
and hence our results will be affected by an unknown systematic error in any event. 
Analysis of
the contributions to $B_\varepsilon(T)$~\cite{Garberoglio20:Beps} shows
that exchange effects are present only for helium isotopes when $T \lesssim
5$~K, hence we will limit ourselves to temperatures higher than that in the
present paper.

Since $\alpha_\mathrm{aniso}(r) = O(r^{-3})$ at large distances, we
observed a slow convergence of the integral leading to $C_\varepsilon(T)$
as a function of the cutoff $R$, in the form
\begin{equation}
  C_\varepsilon(T; R) = C_\varepsilon(T) + \frac{a}{R}.
\label{eq:C_conv}
\end{equation}
This required us to use $R =100$~nm when evaluating the third dielectric
virial coefficient using the superposition approximation of $\mbalpha_3$.
In this case, the asymptotic value obtained by fitting $C_\varepsilon(T; R)$
with a function of the form (\ref{eq:C_conv}) falls within the statistical
uncertainty of the path-integral calculations. 

As usual with path-integral calculations, one has to choose a sufficiently large
value for the Trotter index $P$; in general the optimal value depends on
temperature, as well as the required uncertainty.  We have used the same
values discussed in Ref.~\onlinecite{Garberoglio20:Beps}, namely
$P = \mathrm{nint}(1600~\mathrm{K}/T + 7)$ for ${}^4$He, $P = \mathrm{nint}(
800~\mathrm{K}/T + 4)$ for ${}^{20}$Ne, and $P = \mathrm{nint}(300~\mathrm{K}/T +
4)$ for ${}^{40}$Ar, where $\mathrm{nint}(x)$ denotes the nearest integer to $x$.
The integrals have been evaluated with the parallel implementation of the
VEGAS algorithm.~\cite{pvegas} We found it useful to evaluate separately
the contribution to $C_\varepsilon(T)$ coming from the two-body potential
and polarizability and the contribution due to the three-body potential,
polarizability, and dipole moment. The former converges rather quickly and a
relatively small cutoff $R=6$~nm was sufficient; we used
2~000~000 Monte Carlo samples, and estimated the statistical uncertainty by performing 16
independent runs at each temperature. The second contribution required more
computational effort to produce a reasonably small variance; in this case
we used 8~000~000 Monte Carlo samples and 128 independent runs for each of
the temperatures considered in this work.

We also evaluated the contribution to the uncertainty of our results
obtained by propagating the uncertainties of the potentials and
polarizabilities, where available; in this case we have assumed that the
provided uncertainties are expanded uncertainties at coverage level
$k=2$. Given the exploratory nature of this work, we have used the
straightforward approach of evaluating the third dielectric virial coefficient in the
classical approximation for the perturbed potential or polarizability and
evaluating the standard uncertainty as $1/4$ of the absolute value of the
difference.

\section{Results and Discussion}\label{results}

Before presenting our calculated results and comparing them to experimental data, we emphasize two caveats that apply to the results for all three gases examined.
First, all of the results use the superposition approximation for the three-body polarizability as described in Sec.~\ref{sec:superpos} and the approximation for the three-body dipole moment from Li and Hunt.\cite{Li97}
While these approximations should be accurate in the limit of large interatomic distances, they may not be accurate at shorter distances, and any inaccuracy would produce a systematic error in $C_\varepsilon$.
Second, and related, the use of the superposition approximation and the approximation for the three-body dipole prevents us from making quantitative uncertainty estimates, because of the unknown systematic error.  We have computed and tabulated uncertainties attributable to known factors (uncertainty in the pair and three-body potential and in the pair polarizability, together with statistical uncertainty in the PIMC calculations), but these numbers do not represent a complete uncertainty budget and should not be used in metrological uncertainty budgets.

Because of this unknown systematic uncertainty, the purpose of comparisons with experimental data in this section is not to quantitatively assess the data, but instead to show qualitatively that the present calculations are generally consistent with the limited experimental data.

\subsection{Helium}

In performing the calculations for helium isotopes, we used the pair
potential developed by Czachorowski et al.,~\cite{u2_2020} the three-body
potential by Cencek et al.,~\cite{FCI} and the pair polarizability by
Cencek et al.~\cite{Cencek11} 
All of these quantities have assigned uncertainties. 
The three-body polarizability has been evaluated using the
superposition approximation and the three-body dipole moment has been evaluated with the formulation of Li and Hunt.\cite{Li97} 
At all temperatures, the statistical
uncertainty from the path-integral calculation dominates the uncertainty budget (not counting the unknown uncertainty from the superposition calculation and three-body dipole estimate).

Our results are reported in Tab.~\ref{tab:Ceps_He}. The
values of the third dielectric virial coefficient are negative and their
magnitude decreases with decreasing temperature. The result that we obtain
at $T=300$~K, $C_\varepsilon(300~\mathrm{K}) =
-0.556$~cm${}^9$~mol${}^{-3}$,  is comparable to the only other theoretical result reported in the
literature, from a classical calculation (at ``room temperature'' which should be near 300~K)
by Heller and Gelbart~\cite{Heller74} that used a superposition approximation in combination with relatively simple models for the pair potential and pair polarizability.
Their calculation resulted in $C_\varepsilon = -0.719$~cm${}^9$~mol${}^{-3}$.

The effect of the three-body polarizability in determining the value of
$C_\varepsilon$ for helium is not negligible. As an example, we note
that the third dielectric virial coefficient at $T=300$~K becomes
$-0.441$~cm${}^9$~mol${}^{-3}$ if the three-body polarizability is neglected;
this is roughly a 20\% difference.
In contrast, the effect of the three-body dipole $\mbm_3$ as computed by the approximation of Li and Hunt\cite{Li97} is completely negligible within the precision of our calculations.

In the uncertainty budget, the contributions from the
uncertainty in the three-body potential and in the pair polarizability are of similar size;
that from the pair potential is negligible.
In order to emphasize that the uncertainty in Table~\ref{tab:Ceps_He} and subsequent tables is incomplete due to the unknown systematic error in the superposition approximation, we use the symbol $U^*$ for the expanded ($k = 2$) value rather than the symbol $U$ that would be used for a complete expanded uncertainty.

\begin{table}
  \caption{The third dielectric virial coefficient $C_\varepsilon$ for ${}^4$He and its
    uncertainty. The $U^*(C_\varepsilon)$ are incomplete expanded uncertainties at $k=2$ and
    include the statistical uncertainty of the calculation as well as the
    propagated uncertainties from potentials and the two-body polarizability. There is
    an unknown systematic uncertainty contribution from the superposition
    approximation and the approximation of the three-body dipole term.
  The last column reports the value of the third dielectric virial
  coefficient calculated in the classical approximation.}
  \begin{tabular}{d|d|d|d}
    \multicolumn{1}{c|}{Temperature} &
    \multicolumn{1}{c|}{$C_\varepsilon(T)$} &
    \multicolumn{1}{c|}{$U^*(C_\varepsilon)$} &
    \multicolumn{1}{c}{$C_\varepsilon^\mathrm{cl}(T)$} \\
    \multicolumn{1}{c|}{(K)} &
    \multicolumn{1}{c|}{(cm${}^9$~mol${}^{-3}$)} &
    \multicolumn{1}{c|}{(cm${}^9$~mol${}^{-3}$)} &
    \multicolumn{1}{c}{(cm${}^9$~mol${}^{-3}$)} \\    
    \hline
5	&	-0.251	&	0.089	&	-14.575	\\
7	&	-0.220	&	0.016	&	-1.978	\\
10	&	-0.210	&	0.005	&	-0.413	\\
15	&	-0.214	&	0.003	&	-0.198	\\
20	&	-0.232	&	0.002	&	-0.195	\\
30	&	-0.261	&	0.002	&	-0.228	\\
40	&	-0.287	&	0.001	&	-0.260	\\
50	&	-0.311	&	0.001	&	-0.288	\\
75	&	-0.357	&	0.001	&	-0.342	\\
100	&	-0.398	&	0.001	&	-0.383	\\
125	&	-0.428	&	0.001	&	-0.416	\\
150	&	-0.453	&	0.001	&	-0.444	\\
175	&	-0.476	&	0.001	&	-0.468	\\
200	&	-0.496	&	0.001	&	-0.488	\\
250	&	-0.531	&	0.001	&	-0.523	\\
273.16	&	-0.543	&	0.002	&	-0.537	\\
300	&	-0.556	&	0.001	&	-0.552	\\
350	&	-0.581	&	0.001	&	-0.575	\\
400	&	-0.600	&	0.002	&	-0.595	\\
450	&	-0.617	&	0.002	&	-0.612	\\
500	&	-0.632	&	0.002	&	-0.628	\\
600	&	-0.655	&	0.002	&	-0.652	\\
700	&	-0.676	&	0.002	&	-0.672	\\
800	&	-0.691	&	0.002	&	-0.688	\\
900	&	-0.702	&	0.002	&	-0.701	\\
1000	&	-0.714	&	0.002	&	-0.712	\\
1500	&	-0.745	&	0.002	&	-0.744	\\
2000	&	-0.755	&	0.002	&	-0.755	
\end{tabular}
  \label{tab:Ceps_He}  
\end{table}

\begin{figure}
\includegraphics[width=0.9\linewidth]{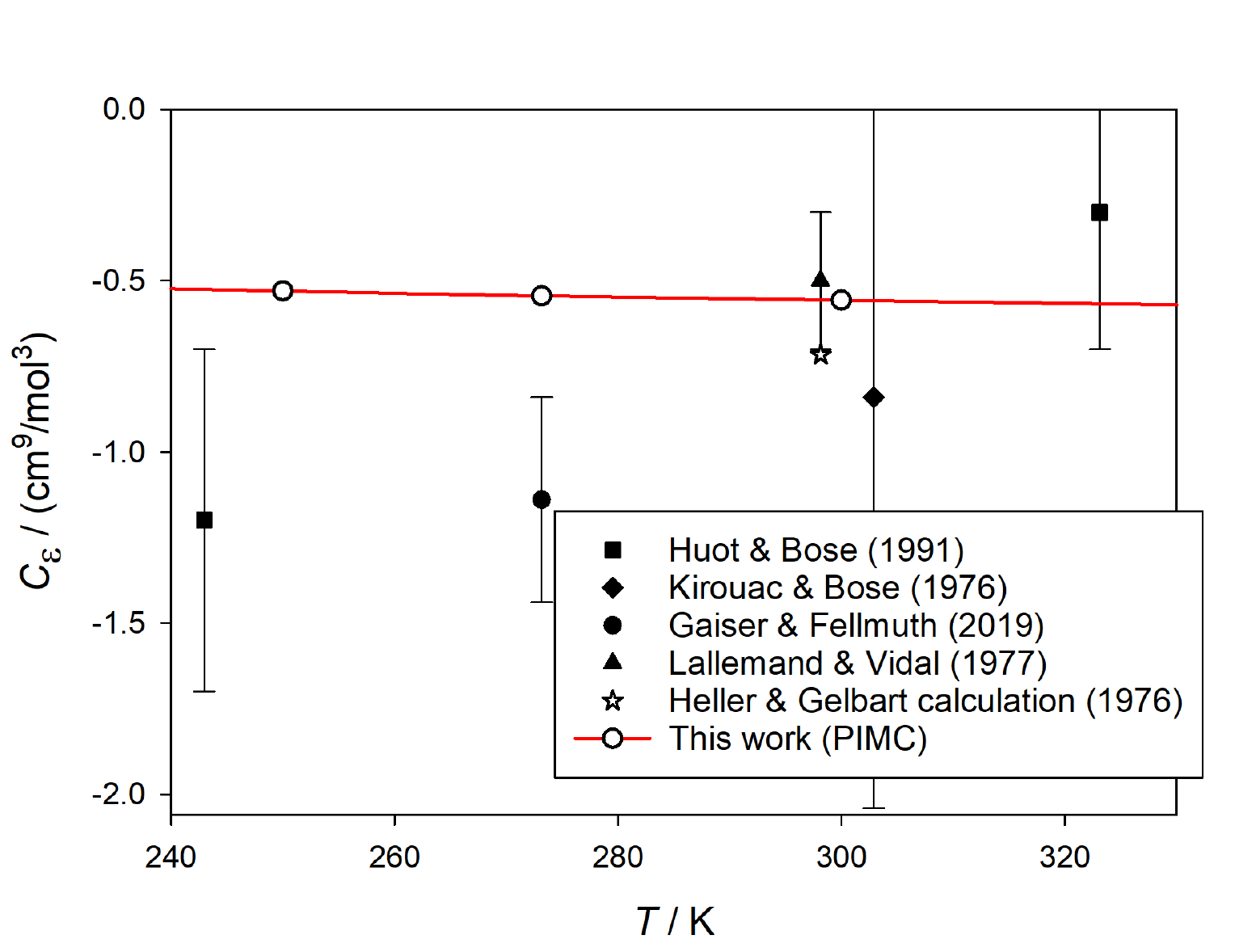}  
  \caption{Comparison of calculated values of $C_\varepsilon(T)$ for ${}^4$He
    with those derived from experiment \cite{Huot91,Kirouac76,Gaiser_2019,Lallemand77} at higher temperatures. The classical calculation of Heller and Gelbart\cite{Heller74} is also shown.}
\label{fig:He-high}  
\end{figure}

In Fig.~\ref{fig:He-high}, our calculations are compared to values of $C_\varepsilon(T)$ found in the literature.
Error bars drawn on the experimental values are those reported in the original publication; in most cases the statistical meaning of the error interval was not stated.
The point of Gaiser and Fellmuth\cite{Gaiser_2019} comes from dielectric-constant gas thermometry experiments where the quantity obtained was a combination of the second and third dielectric and density virial coefficients; this was converted to $C_\varepsilon$ with the use of precise literature values for the third density virial coefficient\cite{Garberoglio2011} and the second dielectric virial coefficient.\cite{Garberoglio20:Beps}
Our results are generally consistent with the experimental data within their considerable scatter, although there is less temperature dependence than might be assumed based on the experimental points shown.

\begin{figure}
\includegraphics[width=0.9\linewidth]{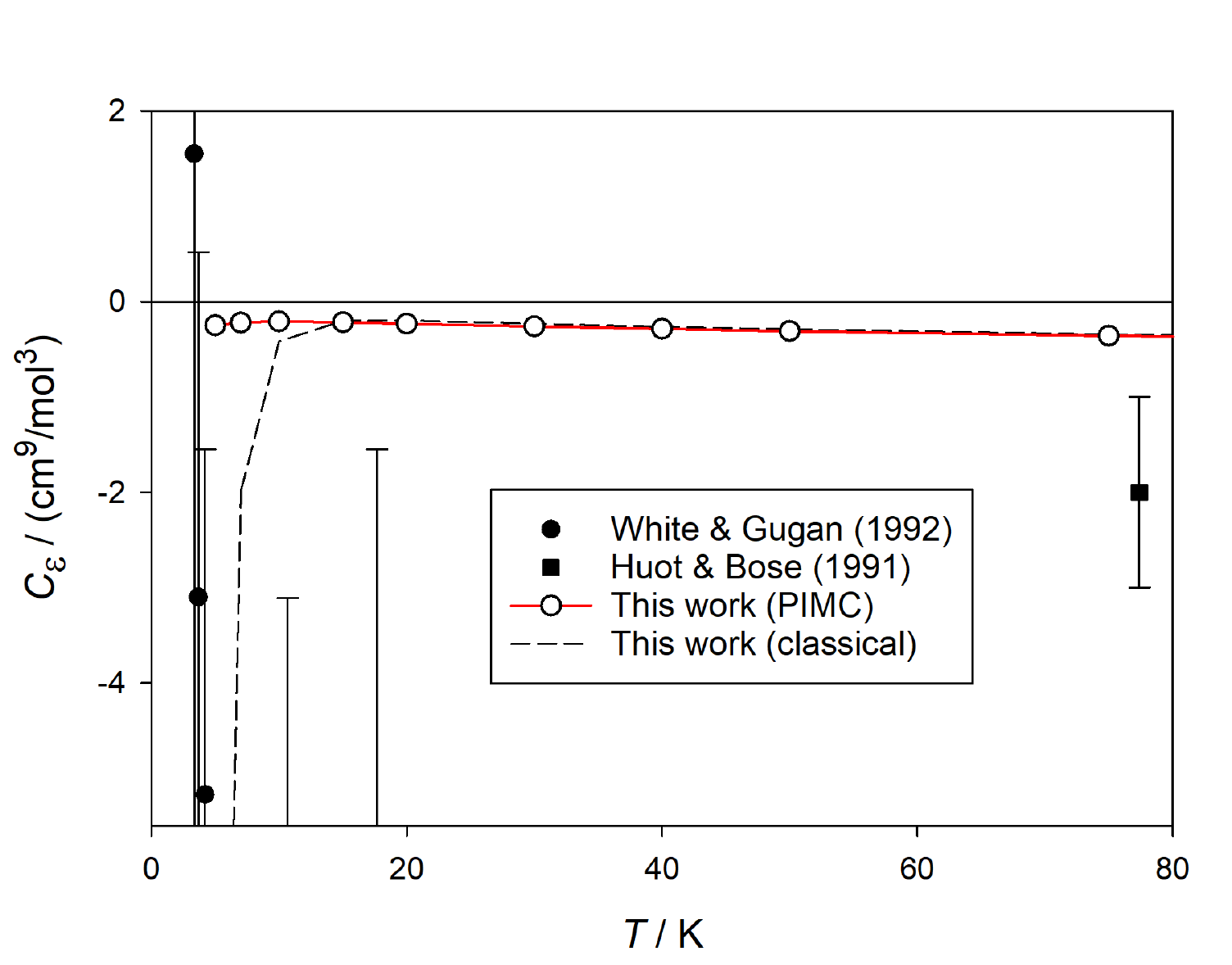}  
  \caption{Comparison of calculated values of $C_\varepsilon(T)$ for ${}^4$He
    with those derived from experiment \cite{Huot91,White_1992} at low temperatures. Two points from White and Gugan\cite{White_1992} are below the bottom of the plot; only the tops of their error bars are visible.}
\label{fig:He-low}  
\end{figure}

Figure~\ref{fig:He-low} shows a comparison with data at low temperatures; note that points by White and Gugan\cite{White_1992} near 11~K and 18~K lie below the bottom of the plot but the tops of their large error bars are visible. 
Our results indicate a relatively flat temperature dependence; it remains to be seen whether this will still be true when a quantitatively accurate three-body polarizability becomes available.
Figure~\ref{fig:He-low} also displays the result of a classical calculation of $C_\varepsilon(T)$, which becomes increasingly inaccurate below 20~K.

\subsection{Neon}

In the case of neon, we used the latest pair potential and pair
polarizability  by Hellmann et al.~\cite{Hellmann21} as well as
the extended Axilrod--Teller three-body potential by Schwerdtfeger and
Hermann.~\cite{Schwerdtfeger09} Dr. Hellmann provided us with the
anisotropic component of the pair polarizability, which was not reported in
the original paper.~\cite{Hellmann_pc}

Our calculated values are reported in Tab.~\ref{tab:Ceps_Ne}.
In this case, we did not perform the propagation of the uncertainty from
the potentials and the two-body polarizability because some of these quantities do not have
an uncertainty estimate, so we report only the statistical uncertainty of our calculation.

It is interesting to note that the effect of the
three-body polarizability on $C_\varepsilon$ is quite significant. Using
classical calculations as an example, the value of $C_\varepsilon$ obtained
neglecting the three-body polarizability at $T=300$~K would be
$-0.914$~cm${}^9$~mol${}^{-3}$, a 50\% difference from the value obtained
including it.
The effect of the approximate three-body dipole term is larger than for $^4$He but still negligible, making $C_\varepsilon$ less negative by an amount on the order of $0.01$~cm${}^9$~mol${}^{-3}$ at temperatures near 300~K.

\begin{table}
  \caption{The third dielectric virial coefficient $C_\varepsilon$ for ${}^{20}$Ne and its
    uncertainty. The $U^*(C_\varepsilon)$ are incomplete expanded uncertainties at $k=2$ and
    only include the statistical uncertainty of the path-integral Monte
    Carlo calculation. 
    Uncertainty contributions from the two- and three-body potentials and polarizabilities
    and three-body dipole moment are not included; in several cases the information needed to 
    estimate these uncertainties is not available.
    The last column reports the value of the third dielectric virial
    coefficient calculated in the classical approximation.}
  \begin{tabular}{d|d|d|d}
    \multicolumn{1}{c|}{Temperature} &
    \multicolumn{1}{c|}{$C_\varepsilon(T)$} &
    \multicolumn{1}{c|}{$U^*(C_\varepsilon)$} &
    \multicolumn{1}{c}{$C_\varepsilon^\mathrm{cl}(T)$} \\
    \multicolumn{1}{c|}{(K)} &
    \multicolumn{1}{c|}{(cm${}^9$~mol${}^{-3}$)} &
    \multicolumn{1}{c|}{(cm${}^9$~mol${}^{-3}$)} &
    \multicolumn{1}{c}{(cm${}^9$~mol${}^{-3}$)} \\    
    \hline
20	&	-19.997	&	0.040	&	-36.961	\\
25	&	-6.167	&	0.026	&	-9.475	\\
30	&	-2.836	&	0.020	&	-3.752	\\
35	&	-1.756	&	0.018	&	-2.063	\\
40	&	-1.332	&	0.016	&	-1.456	\\
45	&	-1.182	&	0.015	&	-1.215	\\
50	&	-1.117	&	0.014	&	-1.117	\\
55	&	-1.088	&	0.012	&	-1.081	\\
60	&	-1.075	&	0.013	&	-1.075	\\
65	&	-1.097	&	0.013	&	-1.084	\\
70	&	-1.113	&	0.012	&	-1.099	\\
75	&	-1.123	&	0.010	&	-1.118	\\
80	&	-1.158	&	0.012	&	-1.139	\\
85	&	-1.174	&	0.011	&	-1.161	\\
90	&	-1.188	&	0.011	&	-1.182	\\
95	&	-1.223	&	0.012	&	-1.204	\\
100	&	-1.236	&	0.011	&	-1.224	\\
150	&	-1.409	&	0.009	&	-1.403	\\
200	&	-1.540	&	0.008	&	-1.540	\\
250	&	-1.647	&	0.008	&	-1.650	\\
300	&	-1.744	&	0.009	&	-1.743	\\
350	&	-1.824	&	0.010	&	-1.822	\\
400	&	-1.890	&	0.009	&	-1.892	\\
500	&	-2.004	&	0.009	&	-2.009	\\
600	&	-2.105	&	0.009	&	-2.105	\\
700	&	-2.183	&	0.009	&	-2.185	\\
800	&	-2.254	&	0.009	&	-2.253	\\
900	&	-2.312	&	0.009	&	-2.312	\\
1000	&	-2.370	&	0.009	&	-2.364	\\
1500	&	-2.555	&	0.009	&	-2.549	\\
2000	&	-2.670	&	0.010	&	-2.662	
  \end{tabular}
  \label{tab:Ceps_Ne}  
\end{table}

\begin{figure}
\includegraphics[width=0.9\linewidth]{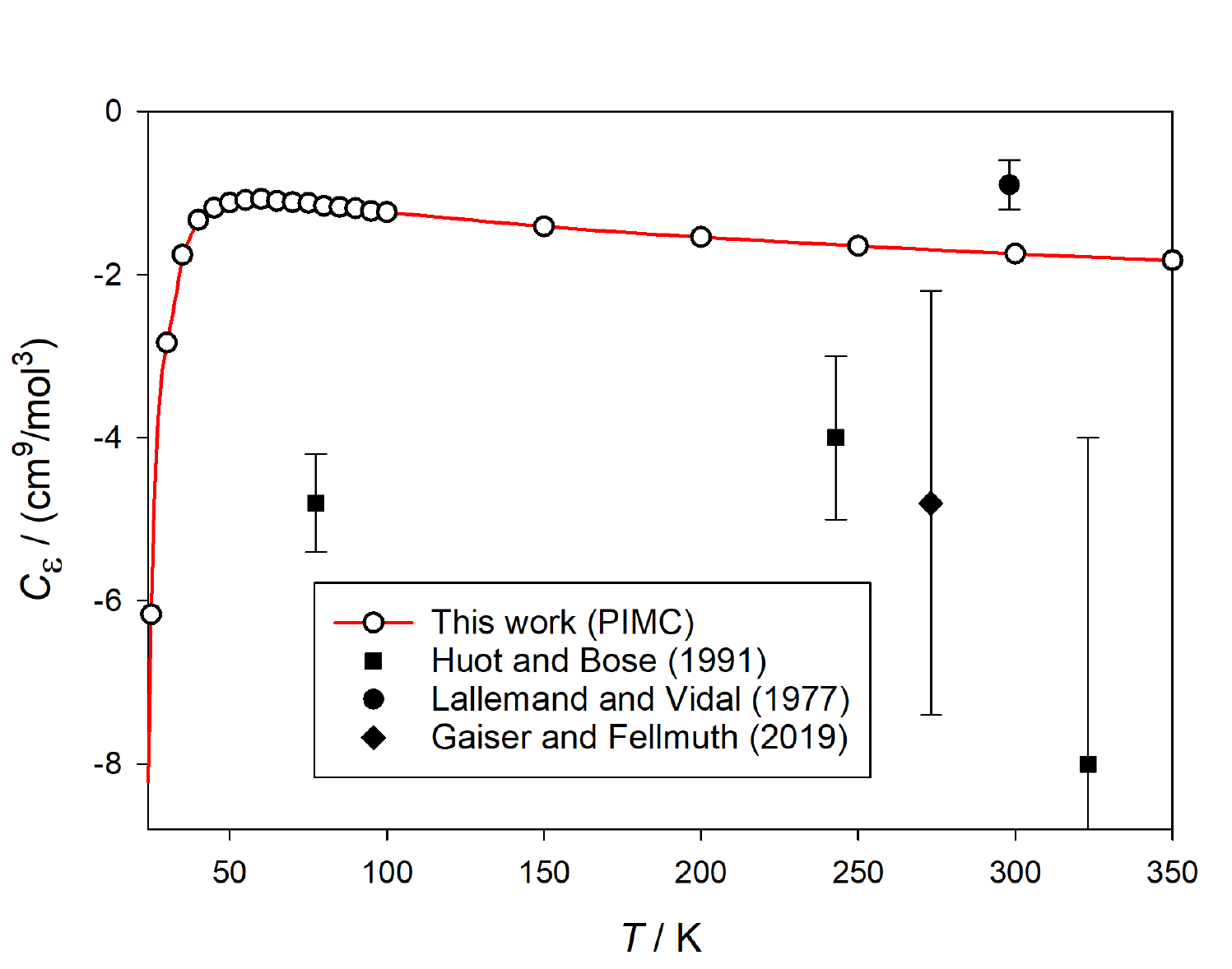}  
  \caption{Comparison of calculated values of $C_\varepsilon(T)$ for neon
    with those derived from experiment. \cite{Huot91,Lallemand77,Gaiser_2019}}
\label{fig:Ne}  
\end{figure}

Figure~\ref{fig:Ne} compares our calculated values with the very limited,
and  mutually inconsistent, experimental data available.
The point from Gaiser and Fellmuth~\cite{Gaiser_2019} is shown with error bars
corresponding to one standard uncertainty and was extracted from their
dielectric-constant gas thermometry data by Rourke.~\cite{Rourke_2021}
Few conclusions can be drawn, especially since we do not know the accuracy
of the three-body approximations, but the large negative values attained
below 40~K might have consequences for gas-based metrology in that
temperature range.

\subsection{Argon}

In the case of ${}^{40}$Ar, we used the pair potential developed by
Patkowski and Szalewicz,~\cite{Patkowski10} the pair polarizability by
Vogel et al.,~\cite{Vogel10} and the three-body potential by Cencek et
al.~\cite{Cencek2013} 
We again estimated the three-body polarizability by the superposition approximation and the three-body dipole moment with the formulation of Li and Hunt.\cite{Li97}
Our results are reported in Tab.~\ref{tab:Ceps_Ar}.

\begin{table}
  \caption{The third dielectric virial coefficient $C_\varepsilon$ for ${}^{40}$Ar and its
    uncertainty. The $U^*(C_\varepsilon)$ are incomplete expanded uncertainties at $k=2$ and
    include the statistical uncertainty of the path-integral Monte
    Carlo calculation and the propagated uncertainty from the potentials and
    the two-body polarizability. There is
    an unknown systematic uncertainty contribution from the superposition
    approximation and the approximation of the three-body dipole term.
    The last column reports the value of the third dielectric virial
    coefficient calculated in the classical approximation.}
  \begin{tabular}{d|d|d|d}
    \multicolumn{1}{c|}{Temperature} &
    \multicolumn{1}{c|}{$C_\varepsilon(T)$} &
    \multicolumn{1}{c|}{$U^*(C_\varepsilon)$} &
    \multicolumn{1}{c}{$C_\varepsilon^\mathrm{cl}(T)$} \\
    \multicolumn{1}{c|}{(K)} &
    \multicolumn{1}{c|}{(cm${}^9$~mol${}^{-3}$)} &
    \multicolumn{1}{c|}{(cm${}^9$~mol${}^{-3}$)} &
    \multicolumn{1}{c}{(cm${}^9$~mol${}^{-3}$)} \\    
    \hline
50	&	7315	&	641	&	7844	\\
75	&	446	&	49	&	467	\\
100	&	13	&	14	&	16	\\
125	&	-60	&	9	&	-60	\\
150	&	-78	&	8	&	-78	\\
175	&	-82	&	7	&	-82	\\
200	&	-82	&	7	&	-82	\\
250	&	-81	&	7	&	-80	\\
273.16	&	-79	&	7	&	-79	\\
300	&	-78	&	7	&	-78	\\
350	&	-76	&	7	&	-76	\\
400	&	-75	&	7	&	-75	\\
450	&	-74	&	7	&	-74	\\
500	&	-73	&	7	&	-73	\\
600	&	-72	&	7	&	-72	\\
700	&	-72	&	7	&	-72	\\
800	&	-72	&	7	&	-72	\\
900	&	-71	&	7	&	-72	\\
1000	&	-71	&	6	&	-72	\\
1500	&	-72	&	6	&	-73	\\
2000	&	-74	&	6	&	-74	
  \end{tabular}
  \label{tab:Ceps_Ar}  
\end{table}

\begin{figure}
\includegraphics[width=0.9\linewidth]{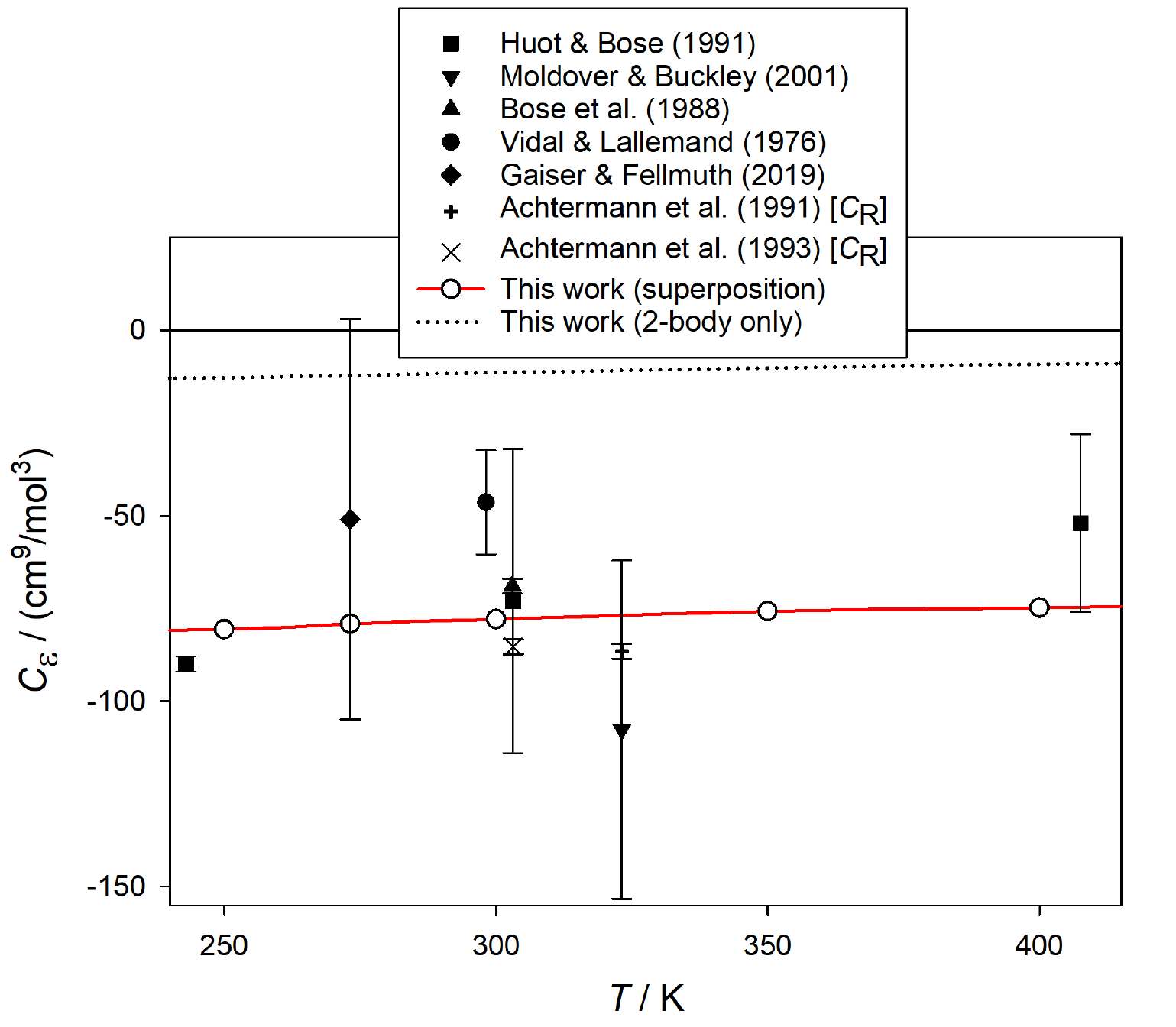}  
  \caption{Comparison of calculated values of $C_\varepsilon(T)$ for argon
    with those derived from experiment. \cite{Bose88,Huot91,Moldover2001,Gaiser_2019,Vidal76,Achtermann91,Achtermann93}}
\label{fig:Ar}  
\end{figure}

Figure~\ref{fig:Ar} shows our calculated results along with experimental values for $C_\varepsilon$, which are somewhat more numerous than for the other two gases.
The Gaiser and Fellmuth point\cite{Gaiser_2019} was obtained in the same manner as described above for helium, using literature values for argon's third density virial coefficient\cite{Cencek2013} and second dielectric virial coefficient.\cite{Garberoglio20:Beps}
The points shown from Achtermann and coworkers\cite{Achtermann91,Achtermann93} are not $C_\varepsilon$ but instead $C_\mathrm{R}$, the third refractivity virial coefficient.  The difference between $C_\varepsilon$ and $C_\mathrm{R}$ is expected to be small ($B_\varepsilon$ and $B_\mathrm{R}$ differ by only about 1\%\cite{Garberoglio20:Beps}), so $C_\mathrm{R}$ still provides a valuable comparison.

Our calculated results are again reasonably consistent with the scattered experimental data.
Figure~\ref{fig:Ar} also shows the results that would be obtained in the absence of the three-body polarizability; it is evident that the three-body polarizability contributes a large amount to $C_\varepsilon$ and is necessary to obtain agreement with experimental data.
In the case of argon, the contribution from the three-body dipole term is not completely negligible; it makes $C_\varepsilon$ less negative by an amount on the order of 1~cm${}^9$~mol${}^{-3}$ at temperatures near 300~K.

\section{Conclusions}
We have presented the first complete framework for calculating the third dielectric virial coefficient of gases with rigorous accounting for quantum effects, including exchange effects.
Calculations of $C_\varepsilon(T)$ were performed for helium, neon, and argon, using the best available pair and three-body potentials and pair polarizability functions.
Our calculations yielded results in qualitative agreement with the limited experimental data available.

The results presented in Sec.~\ref{results} are not yet suitable for rigorous metrological use, because there is an unknown systematic uncertainty due to the use of the superposition approximation for the three-body nonadditive polarizability.
Because the relative contribution of the three-body polarizability to $C_\varepsilon$ is large, especially for argon, an error in the superposition approximation could produce a significant error in $C_\varepsilon(T)$.
Qualitatively, since the superposition approximation produces the correct long-range behavior, we might expect it to be accurate at conditions where dispersion forces dominate the interactions (as is the case for argon at temperatures of practical interest), while perhaps losing accuracy where repulsive configurations dominate the virial coefficients (as is the case for helium at ambient temperatures). 
A possibly analogous situation is the third density virial coefficient of helium, where the Axilrod-Teller three-body potential, which is an induced dipole model valid at long range, produces a correction of the wrong sign above approximately 170~K.\cite{Garberoglio2009b}

There is an additional unknown systematic uncertainty due to the lack of a
three-body dipole moment surface for the gases considered.  While
preliminary calculations using an approximation valid in the long-range
limit suggest that this effect is negligible for helium and neon, and small
for argon, a quantitative estimate is needed for a rigorous uncertainty
budget.  If this three-body dipole contribution is truly small, the
relative uncertainty of the estimate could be large without significantly
increasing the total uncertainty in the calculated $C_\varepsilon$.

It is therefore imperative for the use of these dielectric virial
coefficients in metrology that quantitative surfaces, with uncertainty
estimates, be developed for the three-body polarizability (and, with less
urgency, for the three-body dipole moment).  Such an effort is currently
underway for helium.~\cite{Ceps_forth}

\begin{acknowledgments}
B.J. and G.G. acknowledge support from {\em QuantumPascal} project 18SIB04,
which has received funding from the EMPIR programme co-financed by the
Participating States and from the European Union's Horizon 2020 research and
innovation programme.
B.J. thanks also for the support from the National Science Center, Poland,
Project No. 2017/27/B/ST4/02739.
Code development has been performed on the computing cluster KORE at
Fondazione Bruno Kessler. We acknowledge CINECA (Award No. IscraC-THIDIVI)
under the ISCRA initiative for the availability of high performance
computing resources and support.
\end{acknowledgments}

\section*{Data Availability}
The data that support the findings of this study are available within the article.

\bibliography{Ceps-quantum}

\begin{thebibliography}{59}%
\makeatletter
\providecommand \@ifxundefined [1]{%
 \@ifx{#1\undefined}
}%
\providecommand \@ifnum [1]{%
 \ifnum #1\expandafter \@firstoftwo
 \else \expandafter \@secondoftwo
 \fi
}%
\providecommand \@ifx [1]{%
 \ifx #1\expandafter \@firstoftwo
 \else \expandafter \@secondoftwo
 \fi
}%
\providecommand \natexlab [1]{#1}%
\providecommand \enquote  [1]{``#1''}%
\providecommand \bibnamefont  [1]{#1}%
\providecommand \bibfnamefont [1]{#1}%
\providecommand \citenamefont [1]{#1}%
\providecommand \href@noop [0]{\@secondoftwo}%
\providecommand \href [0]{\begingroup \@sanitize@url \@href}%
\providecommand \@href[1]{\@@startlink{#1}\@@href}%
\providecommand \@@href[1]{\endgroup#1\@@endlink}%
\providecommand \@sanitize@url [0]{\catcode `\\12\catcode `\$12\catcode
  `\&12\catcode `\#12\catcode `\^12\catcode `\_12\catcode `\%12\relax}%
\providecommand \@@startlink[1]{}%
\providecommand \@@endlink[0]{}%
\providecommand \url  [0]{\begingroup\@sanitize@url \@url }%
\providecommand \@url [1]{\endgroup\@href {#1}{\urlprefix }}%
\providecommand \urlprefix  [0]{URL }%
\providecommand \Eprint [0]{\href }%
\providecommand \doibase [0]{http://dx.doi.org/}%
\providecommand \selectlanguage [0]{\@gobble}%
\providecommand \bibinfo  [0]{\@secondoftwo}%
\providecommand \bibfield  [0]{\@secondoftwo}%
\providecommand \translation [1]{[#1]}%
\providecommand \BibitemOpen [0]{}%
\providecommand \bibitemStop [0]{}%
\providecommand \bibitemNoStop [0]{.\EOS\space}%
\providecommand \EOS [0]{\spacefactor3000\relax}%
\providecommand \BibitemShut  [1]{\csname bibitem#1\endcsname}%
\let\auto@bib@innerbib\@empty
\bibitem [{\citenamefont {Gaiser}, \citenamefont {Zandt},\ and\ \citenamefont
  {Fellmuth}(2015)}]{Gaiser_2015}%
  \BibitemOpen
  \bibfield  {author} {\bibinfo {author} {\bibfnamefont {C.}~\bibnamefont
  {Gaiser}}, \bibinfo {author} {\bibfnamefont {T.}~\bibnamefont {Zandt}}, \
  and\ \bibinfo {author} {\bibfnamefont {B.}~\bibnamefont {Fellmuth}},\
  }\bibfield  {title} {\enquote {\bibinfo {title} {Dielectric-constant gas
  thermometry},}\ }\href {\doibase 10.1088/0026-1394/52/5/s217} {\bibfield
  {journal} {\bibinfo  {journal} {Metrologia}\ }\textbf {\bibinfo {volume}
  {52}},\ \bibinfo {pages} {S217--S226} (\bibinfo {year} {2015})}\BibitemShut
  {NoStop}%
\bibitem [{\citenamefont {Gaiser}, \citenamefont {Fellmuth},\ and\
  \citenamefont {Haft}(2017)}]{Gaiser_2017}%
  \BibitemOpen
  \bibfield  {author} {\bibinfo {author} {\bibfnamefont {C.}~\bibnamefont
  {Gaiser}}, \bibinfo {author} {\bibfnamefont {B.}~\bibnamefont {Fellmuth}}, \
  and\ \bibinfo {author} {\bibfnamefont {N.}~\bibnamefont {Haft}},\ }\bibfield
  {title} {\enquote {\bibinfo {title} {Primary thermometry from
  2.5{\hspace{0.167em}}{K} to 140{\hspace{0.167em}}{K} applying
  dielectric-constant gas thermometry},}\ }\href {\doibase
  10.1088/1681-7575/aa5389} {\bibfield  {journal} {\bibinfo  {journal}
  {Metrologia}\ }\textbf {\bibinfo {volume} {54}},\ \bibinfo {pages} {141--147}
  (\bibinfo {year} {2017})}\BibitemShut {NoStop}%
\bibitem [{\citenamefont {Gaiser}\ \emph {et~al.}(2017)\citenamefont {Gaiser},
  \citenamefont {Fellmuth}, \citenamefont {Haft}, \citenamefont {Kuhn},
  \citenamefont {Thiele-Krivoi}, \citenamefont {Zandt}, \citenamefont
  {Fischer}, \citenamefont {Jusko},\ and\ \citenamefont
  {Sabuga}}]{Gaiser_2017b}%
  \BibitemOpen
  \bibfield  {author} {\bibinfo {author} {\bibfnamefont {C.}~\bibnamefont
  {Gaiser}}, \bibinfo {author} {\bibfnamefont {B.}~\bibnamefont {Fellmuth}},
  \bibinfo {author} {\bibfnamefont {N.}~\bibnamefont {Haft}}, \bibinfo {author}
  {\bibfnamefont {A.}~\bibnamefont {Kuhn}}, \bibinfo {author} {\bibfnamefont
  {B.}~\bibnamefont {Thiele-Krivoi}}, \bibinfo {author} {\bibfnamefont
  {T.}~\bibnamefont {Zandt}}, \bibinfo {author} {\bibfnamefont
  {J.}~\bibnamefont {Fischer}}, \bibinfo {author} {\bibfnamefont
  {O.}~\bibnamefont {Jusko}}, \ and\ \bibinfo {author} {\bibfnamefont
  {W.}~\bibnamefont {Sabuga}},\ }\bibfield  {title} {\enquote {\bibinfo {title}
  {Final determination of the {B}oltzmann constant by dielectric-constant gas
  thermometry},}\ }\href {\doibase 10.1088/1681-7575/aa62e3} {\bibfield
  {journal} {\bibinfo  {journal} {Metrologia}\ }\textbf {\bibinfo {volume}
  {54}},\ \bibinfo {pages} {280--289} (\bibinfo {year} {2017})}\BibitemShut
  {NoStop}%
\bibitem [{\citenamefont {Gaiser}, \citenamefont {Fellmuth},\ and\
  \citenamefont {Haft}(2020)}]{Gaiser_2020a}%
  \BibitemOpen
  \bibfield  {author} {\bibinfo {author} {\bibfnamefont {C.}~\bibnamefont
  {Gaiser}}, \bibinfo {author} {\bibfnamefont {B.}~\bibnamefont {Fellmuth}}, \
  and\ \bibinfo {author} {\bibfnamefont {N.}~\bibnamefont {Haft}},\ }\bibfield
  {title} {\enquote {\bibinfo {title} {Thermodynamic-temperature data from 30
  {K} to 200 {K}},}\ }\href {\doibase 10.1088/1681-7575/ab9683} {\bibfield
  {journal} {\bibinfo  {journal} {Metrologia}\ }\textbf {\bibinfo {volume}
  {57}},\ \bibinfo {pages} {055003} (\bibinfo {year} {2020})}\BibitemShut
  {NoStop}%
\bibitem [{\citenamefont {Gaiser}, \citenamefont {Fellmuth},\ and\
  \citenamefont {Sabuga}(2020)}]{Gaiser20}%
  \BibitemOpen
  \bibfield  {author} {\bibinfo {author} {\bibfnamefont {C.}~\bibnamefont
  {Gaiser}}, \bibinfo {author} {\bibfnamefont {B.}~\bibnamefont {Fellmuth}}, \
  and\ \bibinfo {author} {\bibfnamefont {W.}~\bibnamefont {Sabuga}},\
  }\bibfield  {title} {\enquote {\bibinfo {title} {Primary gas-pressure
  standard from electrical measurements and thermophysical ab initio
  calculations.}}\ }\href {\doibase 10.1038/s41567-019-0722-2} {\bibfield
  {journal} {\bibinfo  {journal} {Nature Phys.}\ }\textbf {\bibinfo {volume}
  {16}},\ \bibinfo {pages} {177--180} (\bibinfo {year} {2020})}\BibitemShut
  {NoStop}%
\bibitem [{\citenamefont {Piszczatowski}\ \emph {et~al.}(2015)\citenamefont
  {Piszczatowski}, \citenamefont {Puchalski}, \citenamefont {Komasa},
  \citenamefont {Jeziorski},\ and\ \citenamefont
  {Szalewicz}}]{Piszczatowski15}%
  \BibitemOpen
  \bibfield  {author} {\bibinfo {author} {\bibfnamefont {K.}~\bibnamefont
  {Piszczatowski}}, \bibinfo {author} {\bibfnamefont {M.}~\bibnamefont
  {Puchalski}}, \bibinfo {author} {\bibfnamefont {J.}~\bibnamefont {Komasa}},
  \bibinfo {author} {\bibfnamefont {B.}~\bibnamefont {Jeziorski}}, \ and\
  \bibinfo {author} {\bibfnamefont {K.}~\bibnamefont {Szalewicz}},\ }\bibfield
  {title} {\enquote {\bibinfo {title} {Frequency-dependent polarizability of
  helium including relativistic effects with nuclear recoil terms},}\
  }\href@noop {} {\bibfield  {journal} {\bibinfo  {journal} {Phys. Rev. Lett.}\
  }\textbf {\bibinfo {volume} {114}},\ \bibinfo {pages} {173004} (\bibinfo
  {year} {2015})}\BibitemShut {NoStop}%
\bibitem [{\citenamefont {Rourke}\ \emph {et~al.}(2019)\citenamefont {Rourke},
  \citenamefont {Gaiser}, \citenamefont {Gao}, \citenamefont {{Madonna Ripa}},
  \citenamefont {Moldover}, \citenamefont {Pitre},\ and\ \citenamefont
  {Underwood}}]{Rourke_2019}%
  \BibitemOpen
  \bibfield  {author} {\bibinfo {author} {\bibfnamefont {P.~M.~C.}\
  \bibnamefont {Rourke}}, \bibinfo {author} {\bibfnamefont {C.}~\bibnamefont
  {Gaiser}}, \bibinfo {author} {\bibfnamefont {B.}~\bibnamefont {Gao}},
  \bibinfo {author} {\bibfnamefont {D.}~\bibnamefont {{Madonna Ripa}}},
  \bibinfo {author} {\bibfnamefont {M.~R.}\ \bibnamefont {Moldover}}, \bibinfo
  {author} {\bibfnamefont {L.}~\bibnamefont {Pitre}}, \ and\ \bibinfo {author}
  {\bibfnamefont {R.~J.}\ \bibnamefont {Underwood}},\ }\bibfield  {title}
  {\enquote {\bibinfo {title} {Refractive-index gas thermometry},}\ }\href
  {\doibase 10.1088/1681-7575/ab0dbe} {\bibfield  {journal} {\bibinfo
  {journal} {Metrologia}\ }\textbf {\bibinfo {volume} {56}},\ \bibinfo {pages}
  {032001} (\bibinfo {year} {2019})}\BibitemShut {NoStop}%
\bibitem [{\citenamefont {Gao}\ \emph {et~al.}(2020)\citenamefont {Gao},
  \citenamefont {Zhang}, \citenamefont {Han}, \citenamefont {Pan},
  \citenamefont {Chen}, \citenamefont {Song}, \citenamefont {Liu},
  \citenamefont {Hu}, \citenamefont {Kong}, \citenamefont {Sparasci},
  \citenamefont {Plimmer}, \citenamefont {Luo},\ and\ \citenamefont
  {Pitre}}]{Gao_2020}%
  \BibitemOpen
  \bibfield  {author} {\bibinfo {author} {\bibfnamefont {B.}~\bibnamefont
  {Gao}}, \bibinfo {author} {\bibfnamefont {H.}~\bibnamefont {Zhang}}, \bibinfo
  {author} {\bibfnamefont {D.}~\bibnamefont {Han}}, \bibinfo {author}
  {\bibfnamefont {C.}~\bibnamefont {Pan}}, \bibinfo {author} {\bibfnamefont
  {H.}~\bibnamefont {Chen}}, \bibinfo {author} {\bibfnamefont {Y.}~\bibnamefont
  {Song}}, \bibinfo {author} {\bibfnamefont {W.}~\bibnamefont {Liu}}, \bibinfo
  {author} {\bibfnamefont {J.}~\bibnamefont {Hu}}, \bibinfo {author}
  {\bibfnamefont {X.}~\bibnamefont {Kong}}, \bibinfo {author} {\bibfnamefont
  {F.}~\bibnamefont {Sparasci}}, \bibinfo {author} {\bibfnamefont
  {M.}~\bibnamefont {Plimmer}}, \bibinfo {author} {\bibfnamefont
  {E.}~\bibnamefont {Luo}}, \ and\ \bibinfo {author} {\bibfnamefont
  {L.}~\bibnamefont {Pitre}},\ }\bibfield  {title} {\enquote {\bibinfo {title}
  {Measurement of thermodynamic temperature between 5 {K} and 24.5 {K} with
  single-pressure refractive-index gas thermometry},}\ }\href {\doibase
  10.1088/1681-7575/ab84ca} {\bibfield  {journal} {\bibinfo  {journal}
  {Metrologia}\ }\textbf {\bibinfo {volume} {57}},\ \bibinfo {pages} {065006}
  (\bibinfo {year} {2020})}\BibitemShut {NoStop}%
\bibitem [{\citenamefont {{Madonna Ripa}}\ \emph {et~al.}(2021)\citenamefont
  {{Madonna Ripa}}, \citenamefont {Imbraguglio}, \citenamefont {Gaiser},
  \citenamefont {Steur}, \citenamefont {Giraudi}, \citenamefont {Fogliati},
  \citenamefont {Bertinetti}, \citenamefont {Lopardo}, \citenamefont
  {Dematteis},\ and\ \citenamefont {Gavioso}}]{Ripa_2021}%
  \BibitemOpen
  \bibfield  {author} {\bibinfo {author} {\bibfnamefont {D.}~\bibnamefont
  {{Madonna Ripa}}}, \bibinfo {author} {\bibfnamefont {D.}~\bibnamefont
  {Imbraguglio}}, \bibinfo {author} {\bibfnamefont {C.}~\bibnamefont {Gaiser}},
  \bibinfo {author} {\bibfnamefont {P.~P.~M.}\ \bibnamefont {Steur}}, \bibinfo
  {author} {\bibfnamefont {D.}~\bibnamefont {Giraudi}}, \bibinfo {author}
  {\bibfnamefont {M.}~\bibnamefont {Fogliati}}, \bibinfo {author}
  {\bibfnamefont {M.}~\bibnamefont {Bertinetti}}, \bibinfo {author}
  {\bibfnamefont {G.}~\bibnamefont {Lopardo}}, \bibinfo {author} {\bibfnamefont
  {R.}~\bibnamefont {Dematteis}}, \ and\ \bibinfo {author} {\bibfnamefont
  {R.~M.}\ \bibnamefont {Gavioso}},\ }\bibfield  {title} {\enquote {\bibinfo
  {title} {Refractive index gas thermometry between 13.8 {K} and 161.4 {K}},}\
  }\href {\doibase 10.1088/1681-7575/abe249} {\bibfield  {journal} {\bibinfo
  {journal} {Metrologia}\ }\textbf {\bibinfo {volume} {58}},\ \bibinfo {pages}
  {025008} (\bibinfo {year} {2021})}\BibitemShut {NoStop}%
\bibitem [{\citenamefont {Rourke}(2021)}]{Rourke_2021}%
  \BibitemOpen
  \bibfield  {author} {\bibinfo {author} {\bibfnamefont {P.~M.}\ \bibnamefont
  {Rourke}},\ }\bibfield  {title} {\enquote {\bibinfo {title} {Perspective on
  the refractive-index gas metrology data landscape},}\ }\href@noop {}
  {\bibfield  {journal} {\bibinfo  {journal} {J. Phys. Chem. Ref. Data}\
  }\textbf {\bibinfo {volume} {50}},\ \bibinfo {pages} {033104} (\bibinfo
  {year} {2021})}\BibitemShut {NoStop}%
\bibitem [{\citenamefont {Garberoglio}\ and\ \citenamefont
  {Harvey}(2020{\natexlab{a}})}]{Garberoglio20:Beps}%
  \BibitemOpen
  \bibfield  {author} {\bibinfo {author} {\bibfnamefont {G.}~\bibnamefont
  {Garberoglio}}\ and\ \bibinfo {author} {\bibfnamefont {A.~H.}\ \bibnamefont
  {Harvey}},\ }\bibfield  {title} {\enquote {\bibinfo {title} {Path-integral
  calculation of the second dielectric and refractivity virial coefficients of
  helium, neon, and argon},}\ }\href@noop {} {\bibfield  {journal} {\bibinfo
  {journal} {J. Res. Natl. Inst. Stand. Technol.}\ }\textbf {\bibinfo {volume}
  {125}},\ \bibinfo {pages} {125022} (\bibinfo {year}
  {2020}{\natexlab{a}})}\BibitemShut {NoStop}%
\bibitem [{\citenamefont {Puchalski}\ \emph {et~al.}(2020)\citenamefont
  {Puchalski}, \citenamefont {Szalewicz}, \citenamefont {Lesiuk},\ and\
  \citenamefont {Jeziorski}}]{Puchalski20}%
  \BibitemOpen
  \bibfield  {author} {\bibinfo {author} {\bibfnamefont {M.}~\bibnamefont
  {Puchalski}}, \bibinfo {author} {\bibfnamefont {K.}~\bibnamefont
  {Szalewicz}}, \bibinfo {author} {\bibfnamefont {M.}~\bibnamefont {Lesiuk}}, \
  and\ \bibinfo {author} {\bibfnamefont {B.}~\bibnamefont {Jeziorski}},\
  }\bibfield  {title} {\enquote {\bibinfo {title} {{QED} calculation of the
  dipole polarizability of helium atom},}\ }\href {\doibase
  10.1103/PhysRevA.101.022505} {\bibfield  {journal} {\bibinfo  {journal}
  {Phys. Rev. A}\ }\textbf {\bibinfo {volume} {101}},\ \bibinfo {pages}
  {022505} (\bibinfo {year} {2020})}\BibitemShut {NoStop}%
\bibitem [{\citenamefont {Hirschfelder}, \citenamefont {Curtiss},\ and\
  \citenamefont {Bird}(1954)}]{Hirschfelder:54}%
  \BibitemOpen
  \bibfield  {author} {\bibinfo {author} {\bibfnamefont {J.~O.}\ \bibnamefont
  {Hirschfelder}}, \bibinfo {author} {\bibfnamefont {C.~F.}\ \bibnamefont
  {Curtiss}}, \ and\ \bibinfo {author} {\bibfnamefont {R.~B.}\ \bibnamefont
  {Bird}},\ }\href@noop {} {\emph {\bibinfo {title} {Molecular Theory of Gases
  and Liquids}}}\ (\bibinfo  {publisher} {John Wiley \& Sons},\ \bibinfo
  {address} {New York},\ \bibinfo {year} {1954})\BibitemShut {NoStop}%
\bibitem [{\citenamefont {Song}\ and\ \citenamefont {Luo}(2020)}]{Song_2020}%
  \BibitemOpen
  \bibfield  {author} {\bibinfo {author} {\bibfnamefont {B.}~\bibnamefont
  {Song}}\ and\ \bibinfo {author} {\bibfnamefont {Q.-Y.}\ \bibnamefont {Luo}},\
  }\bibfield  {title} {\enquote {\bibinfo {title} {Accurate second dielectric
  virial coefficient of helium, neon, and argon from {\em ab initio} potentials
  and polarizabilities},}\ }\href {\doibase 10.1088/1681-7575/ab62c3}
  {\bibfield  {journal} {\bibinfo  {journal} {Metrologia}\ }\textbf {\bibinfo
  {volume} {57}},\ \bibinfo {pages} {025007} (\bibinfo {year}
  {2020})}\BibitemShut {NoStop}%
\bibitem [{\citenamefont {Garberoglio}\ and\ \citenamefont
  {Harvey}(2009)}]{Garberoglio2009b}%
  \BibitemOpen
  \bibfield  {author} {\bibinfo {author} {\bibfnamefont {G.}~\bibnamefont
  {Garberoglio}}\ and\ \bibinfo {author} {\bibfnamefont {A.~H.}\ \bibnamefont
  {Harvey}},\ }\bibfield  {title} {\enquote {\bibinfo {title} {First-principles
  calculation of the third virial coefficient of helium},}\ }\href {\doibase
  10.6028/jres.114.018} {\bibfield  {journal} {\bibinfo  {journal} {J. Res.
  Nat. Inst. Stand. Technol.}\ }\textbf {\bibinfo {volume} {114}},\ \bibinfo
  {pages} {249--262} (\bibinfo {year} {2009})}\BibitemShut {NoStop}%
\bibitem [{\citenamefont {Garberoglio}, \citenamefont {Moldover},\ and\
  \citenamefont {Harvey}(2011)}]{Garberoglio2011}%
  \BibitemOpen
  \bibfield  {author} {\bibinfo {author} {\bibfnamefont {G.}~\bibnamefont
  {Garberoglio}}, \bibinfo {author} {\bibfnamefont {M.~R.}\ \bibnamefont
  {Moldover}}, \ and\ \bibinfo {author} {\bibfnamefont {A.~H.}\ \bibnamefont
  {Harvey}},\ }\bibfield  {title} {\enquote {\bibinfo {title} {Improved
  first-principles calculation of the third virial coefficient of helium},}\
  }\href {\doibase 10.6028/jres.116.016} {\bibfield  {journal} {\bibinfo
  {journal} {J. Res. Nat. Inst. Stand. Technol.}\ }\textbf {\bibinfo {volume}
  {116}},\ \bibinfo {pages} {729--742} (\bibinfo {year} {2011})}\BibitemShut
  {NoStop}%
\bibitem [{\citenamefont {Heller}\ and\ \citenamefont
  {Gelbart}(1974)}]{Heller74}%
  \BibitemOpen
  \bibfield  {author} {\bibinfo {author} {\bibfnamefont {D.~F.}\ \bibnamefont
  {Heller}}\ and\ \bibinfo {author} {\bibfnamefont {W.~M.}\ \bibnamefont
  {Gelbart}},\ }\bibfield  {title} {\enquote {\bibinfo {title} {Short range
  electronic distortion and the density dependent dielectric function of simple
  gases},}\ }\href {\doibase https://doi.org/10.1016/0009-2614(74)90241-3}
  {\bibfield  {journal} {\bibinfo  {journal} {Chem. Phys. Lett.}\ }\textbf
  {\bibinfo {volume} {27}},\ \bibinfo {pages} {359--364} (\bibinfo {year}
  {1974})}\BibitemShut {NoStop}%
\bibitem [{\citenamefont {Puchalski}\ \emph {et~al.}(2016)\citenamefont
  {Puchalski}, \citenamefont {Piszczatowski}, \citenamefont {Komasa},
  \citenamefont {Jeziorski},\ and\ \citenamefont {Szalewicz}}]{Puchalski16}%
  \BibitemOpen
  \bibfield  {author} {\bibinfo {author} {\bibfnamefont {M.}~\bibnamefont
  {Puchalski}}, \bibinfo {author} {\bibfnamefont {K.}~\bibnamefont
  {Piszczatowski}}, \bibinfo {author} {\bibfnamefont {J.}~\bibnamefont
  {Komasa}}, \bibinfo {author} {\bibfnamefont {B.}~\bibnamefont {Jeziorski}}, \
  and\ \bibinfo {author} {\bibfnamefont {K.}~\bibnamefont {Szalewicz}},\
  }\bibfield  {title} {\enquote {\bibinfo {title} {Theoretical determination of
  the polarizability dispersion and the refractive index of helium},}\ }\href
  {\doibase 10.1103/PhysRevA.93.032515} {\bibfield  {journal} {\bibinfo
  {journal} {Phys. Rev. A}\ }\textbf {\bibinfo {volume} {93}},\ \bibinfo
  {pages} {032515} (\bibinfo {year} {2016})}\BibitemShut {NoStop}%
\bibitem [{\citenamefont {Rourke}(2017)}]{Rourke_2017}%
  \BibitemOpen
  \bibfield  {author} {\bibinfo {author} {\bibfnamefont {P.~M.~C.}\
  \bibnamefont {Rourke}},\ }\bibfield  {title} {\enquote {\bibinfo {title}
  {{NRC} microwave refractive index gas thermometry implementation between 24.5
  {K} and 84 {K}},}\ }\href {\doibase 10.1007/s10765-017-2239-1} {\bibfield
  {journal} {\bibinfo  {journal} {Int. J. Thermophys.}\ }\textbf {\bibinfo
  {volume} {38}},\ \bibinfo {pages} {107} (\bibinfo {year} {2017})}\BibitemShut
  {NoStop}%
\bibitem [{\citenamefont {Gaiser}\ and\ \citenamefont
  {Fellmuth}(2019)}]{Gaiser_2019}%
  \BibitemOpen
  \bibfield  {author} {\bibinfo {author} {\bibfnamefont {C.}~\bibnamefont
  {Gaiser}}\ and\ \bibinfo {author} {\bibfnamefont {B.}~\bibnamefont
  {Fellmuth}},\ }\bibfield  {title} {\enquote {\bibinfo {title}
  {Highly-accurate density-virial-coefficient values for helium, neon, and
  argon at 0.01 \textdegree {C} determined by dielectric-constant gas
  thermometry},}\ }\href {\doibase 10.1063/1.5090224} {\bibfield  {journal}
  {\bibinfo  {journal} {J. Chem. Phys.}\ }\textbf {\bibinfo {volume} {150}},\
  \bibinfo {pages} {134303} (\bibinfo {year} {2019})}\BibitemShut {NoStop}%
\bibitem [{\citenamefont {O'Brien}\ \emph {et~al.}(1973)\citenamefont
  {O'Brien}, \citenamefont {Gutschick}, \citenamefont {McKoy},\ and\
  \citenamefont {McTague}}]{OB_1973}%
  \BibitemOpen
  \bibfield  {author} {\bibinfo {author} {\bibfnamefont {E.~F.}\ \bibnamefont
  {O'Brien}}, \bibinfo {author} {\bibfnamefont {V.~P.}\ \bibnamefont
  {Gutschick}}, \bibinfo {author} {\bibfnamefont {V.}~\bibnamefont {McKoy}}, \
  and\ \bibinfo {author} {\bibfnamefont {J.~P.}\ \bibnamefont {McTague}},\
  }\bibfield  {title} {\enquote {\bibinfo {title} {Polarizability of
  interacting atoms: Relation to collision-induced light scattering and
  dielectric models},}\ }\href {\doibase 10.1103/PhysRevA.8.690} {\bibfield
  {journal} {\bibinfo  {journal} {Phys. Rev. A}\ }\textbf {\bibinfo {volume}
  {8}},\ \bibinfo {pages} {690--696} (\bibinfo {year} {1973})}\BibitemShut
  {NoStop}%
\bibitem [{\citenamefont {Alder}\ \emph {et~al.}(1980)\citenamefont {Alder},
  \citenamefont {Beers}, \citenamefont {Strauss},\ and\ \citenamefont
  {Weis}}]{Alder80}%
  \BibitemOpen
  \bibfield  {author} {\bibinfo {author} {\bibfnamefont {B.~J.}\ \bibnamefont
  {Alder}}, \bibinfo {author} {\bibfnamefont {J.~C.}\ \bibnamefont {Beers}},
  \bibinfo {author} {\bibfnamefont {H.~L.}\ \bibnamefont {Strauss}}, \ and\
  \bibinfo {author} {\bibfnamefont {J.~J.}\ \bibnamefont {Weis}},\ }\bibfield
  {title} {\enquote {\bibinfo {title} {Dielectric constant of atomic fluids
  with variable polarizability},}\ }\href {\doibase 10.1073/pnas.77.6.3098}
  {\bibfield  {journal} {\bibinfo  {journal} {Proc. Nat. Acad. Sci.}\ }\textbf
  {\bibinfo {volume} {77}},\ \bibinfo {pages} {3098--3102} (\bibinfo {year}
  {1980})}\BibitemShut {NoStop}%
\bibitem [{\citenamefont {Vidal}\ and\ \citenamefont
  {Lallemand}(1976)}]{Vidal76}%
  \BibitemOpen
  \bibfield  {author} {\bibinfo {author} {\bibfnamefont {D.}~\bibnamefont
  {Vidal}}\ and\ \bibinfo {author} {\bibfnamefont {M.}~\bibnamefont
  {Lallemand}},\ }\bibfield  {title} {\enquote {\bibinfo {title} {Evolution of
  the {C}lausius--{M}ossotti function of noble gases and nitrogen, at moderate
  and high density, near room temperature},}\ }\href {\doibase
  10.1063/1.432114} {\bibfield  {journal} {\bibinfo  {journal} {J. Chem.
  Phys.}\ }\textbf {\bibinfo {volume} {64}},\ \bibinfo {pages} {4293--4302}
  (\bibinfo {year} {1976})}\BibitemShut {NoStop}%
\bibitem [{\citenamefont {Lallemand}\ and\ \citenamefont
  {Vidal}(1977)}]{Lallemand77}%
  \BibitemOpen
  \bibfield  {author} {\bibinfo {author} {\bibfnamefont {M.}~\bibnamefont
  {Lallemand}}\ and\ \bibinfo {author} {\bibfnamefont {D.}~\bibnamefont
  {Vidal}},\ }\bibfield  {title} {\enquote {\bibinfo {title} {Variation of the
  polarizability of noble gases with density},}\ }\href {\doibase
  10.1063/1.433839} {\bibfield  {journal} {\bibinfo  {journal} {J. Chem.
  Phys.}\ }\textbf {\bibinfo {volume} {66}},\ \bibinfo {pages} {4776--4780}
  (\bibinfo {year} {1977})}\BibitemShut {NoStop}%
\bibitem [{\citenamefont {Hill}(1958)}]{Hill58}%
  \BibitemOpen
  \bibfield  {author} {\bibinfo {author} {\bibfnamefont {T.~L.}\ \bibnamefont
  {Hill}},\ }\bibfield  {title} {\enquote {\bibinfo {title} {Theory of the
  dielectric constant of imperfect gases and dilute solutions},}\ }\href@noop
  {} {\bibfield  {journal} {\bibinfo  {journal} {J. Chem. Phys.}\ }\textbf
  {\bibinfo {volume} {28}},\ \bibinfo {pages} {61--66} (\bibinfo {year}
  {1958})}\BibitemShut {NoStop}%
\bibitem [{\citenamefont {Moszynski}, \citenamefont {Heijmen},\ and\
  \citenamefont {{van der Avoird}}(1995)}]{Mos1}%
  \BibitemOpen
  \bibfield  {author} {\bibinfo {author} {\bibfnamefont {R.}~\bibnamefont
  {Moszynski}}, \bibinfo {author} {\bibfnamefont {T.~G.~A.}\ \bibnamefont
  {Heijmen}}, \ and\ \bibinfo {author} {\bibfnamefont {A.}~\bibnamefont {{van
  der Avoird}}},\ }\bibfield  {title} {\enquote {\bibinfo {title} {Second
  dielectric virial coefficient of helium gas: quantum-statistical calculations
  from an ab initio interaction-induced polarizability},}\ }\href@noop {}
  {\bibfield  {journal} {\bibinfo  {journal} {Chem. Phys. Lett.}\ }\textbf
  {\bibinfo {volume} {247}},\ \bibinfo {pages} {440--446} (\bibinfo {year}
  {1995})}\BibitemShut {NoStop}%
\bibitem [{\citenamefont {Gray}, \citenamefont {Gubbins},\ and\ \citenamefont
  {Joslin}(2011)}]{GG2}%
  \BibitemOpen
  \bibfield  {author} {\bibinfo {author} {\bibfnamefont {C.~G.}\ \bibnamefont
  {Gray}}, \bibinfo {author} {\bibfnamefont {K.~E.}\ \bibnamefont {Gubbins}}, \
  and\ \bibinfo {author} {\bibfnamefont {C.~G.}\ \bibnamefont {Joslin}},\
  }\href@noop {} {\emph {\bibinfo {title} {Theory of Molecular Fluids}}},\
  Vol.\ \bibinfo {volume} {2: {A}pplications}\ (\bibinfo  {publisher} {Oxford
  Science Publications},\ \bibinfo {year} {2011})\BibitemShut {NoStop}%
\bibitem [{\citenamefont {Jackson}(1998)}]{JDJ}%
  \BibitemOpen
  \bibfield  {author} {\bibinfo {author} {\bibfnamefont {J.~D.}\ \bibnamefont
  {Jackson}},\ }\href@noop {} {\emph {\bibinfo {title} {Classical
  Electrodynamics}}},\ \bibinfo {edition} {3rd}\ ed.\ (\bibinfo  {publisher}
  {John Wiley \& Sons},\ \bibinfo {address} {New York},\ \bibinfo {year}
  {1998})\BibitemShut {NoStop}%
\bibitem [{\citenamefont {Garberoglio}\ \emph {et~al.}()\citenamefont
  {Garberoglio}, \citenamefont {Harvey}, \citenamefont {Lang}, \citenamefont
  {Lesiuk}, \citenamefont {Przybytek},\ and\ \citenamefont
  {Jeziorski}}]{Ceps_forth}%
  \BibitemOpen
  \bibfield  {author} {\bibinfo {author} {\bibfnamefont {G.}~\bibnamefont
  {Garberoglio}}, \bibinfo {author} {\bibfnamefont {A.~H.}\ \bibnamefont
  {Harvey}}, \bibinfo {author} {\bibfnamefont {J.}~\bibnamefont {Lang}},
  \bibinfo {author} {\bibfnamefont {M.}~\bibnamefont {Lesiuk}}, \bibinfo
  {author} {\bibfnamefont {M.}~\bibnamefont {Przybytek}}, \ and\ \bibinfo
  {author} {\bibfnamefont {B.}~\bibnamefont {Jeziorski}},\ }\href@noop {} {\
  }\bibinfo {note} {To be published}\BibitemShut {NoStop}%
\bibitem [{\citenamefont {Hill}(1959)}]{Hill59}%
  \BibitemOpen
  \bibfield  {author} {\bibinfo {author} {\bibfnamefont {T.~L.}\ \bibnamefont
  {Hill}},\ }\bibfield  {title} {\enquote {\bibinfo {title} {Theory of the
  dielectric constant of dilute solutions. {II}},}\ }\href@noop {} {\bibfield
  {journal} {\bibinfo  {journal} {J. Chem. Phys.}\ }\textbf {\bibinfo {volume}
  {30}},\ \bibinfo {pages} {1114} (\bibinfo {year} {1959})}\BibitemShut
  {NoStop}%
\bibitem [{\citenamefont {Martin}(1974)}]{Martin74}%
  \BibitemOpen
  \bibfield  {author} {\bibinfo {author} {\bibfnamefont {P.~H.}\ \bibnamefont
  {Martin}},\ }\bibfield  {title} {\enquote {\bibinfo {title} {The long-range
  dipole moment of three identical atoms},}\ }\href@noop {} {\bibfield
  {journal} {\bibinfo  {journal} {Mol. Phys.}\ }\textbf {\bibinfo {volume}
  {27}},\ \bibinfo {pages} {129--134} (\bibinfo {year} {1974})}\BibitemShut
  {NoStop}%
\bibitem [{\citenamefont {Bruch}, \citenamefont {Corcoran},\ and\ \citenamefont
  {Weinhold}(1978)}]{Bruch78}%
  \BibitemOpen
  \bibfield  {author} {\bibinfo {author} {\bibfnamefont {L.~W.}\ \bibnamefont
  {Bruch}}, \bibinfo {author} {\bibfnamefont {C.~T.}\ \bibnamefont {Corcoran}},
  \ and\ \bibinfo {author} {\bibfnamefont {F.}~\bibnamefont {Weinhold}},\
  }\bibfield  {title} {\enquote {\bibinfo {title} {On the dipole moment of
  three identical spherical atoms},}\ }\href@noop {} {\bibfield  {journal}
  {\bibinfo  {journal} {Mol. Phys.}\ }\textbf {\bibinfo {volume} {35}},\
  \bibinfo {pages} {1205--1210} (\bibinfo {year} {1978})}\BibitemShut {NoStop}%
\bibitem [{\citenamefont {Li}\ and\ \citenamefont {Hunt}(1997)}]{Li97}%
  \BibitemOpen
  \bibfield  {author} {\bibinfo {author} {\bibfnamefont {X.}~\bibnamefont
  {Li}}\ and\ \bibinfo {author} {\bibfnamefont {K.~L.~C.}\ \bibnamefont
  {Hunt}},\ }\bibfield  {title} {\enquote {\bibinfo {title} {Nonadditive
  three-body dipoles of inert gas trimers and
  {H$_2\cdots\mathrm{H}_2\cdots\mathrm{H}_2$}: Long-range effects in far
  infrared absorption and triple vibrational transitions},}\ }\href {\doibase
  10.1063/1.474790} {\bibfield  {journal} {\bibinfo  {journal} {J. Chem.
  Phys.}\ }\textbf {\bibinfo {volume} {107}},\ \bibinfo {pages} {4133--4153}
  (\bibinfo {year} {1997})}\BibitemShut {NoStop}%
\bibitem [{\citenamefont {Buckingham}\ and\ \citenamefont
  {Hands}(1991)}]{Buck91}%
  \BibitemOpen
  \bibfield  {author} {\bibinfo {author} {\bibfnamefont {A.}~\bibnamefont
  {Buckingham}}\ and\ \bibinfo {author} {\bibfnamefont {I.}~\bibnamefont
  {Hands}},\ }\bibfield  {title} {\enquote {\bibinfo {title} {The three-body
  contribution to the polarizability of a trimer of inert gas atoms using a
  dipole--induced-dipole model},}\ }\href {\doibase
  https://doi.org/10.1016/0009-2614(91)80257-X} {\bibfield  {journal} {\bibinfo
   {journal} {Chem. Phys. Lett.}\ }\textbf {\bibinfo {volume} {185}},\ \bibinfo
  {pages} {544--549} (\bibinfo {year} {1991})}\BibitemShut {NoStop}%
\bibitem [{\citenamefont {Gelbart}(1974)}]{Gelbart74}%
  \BibitemOpen
  \bibfield  {author} {\bibinfo {author} {\bibfnamefont {W.~M.}\ \bibnamefont
  {Gelbart}},\ }\enquote {\bibinfo {title} {Depolarized light scattering by
  simple fluids},}\ in\ \href {\doibase 10.1002/9780470143780.ch1} {\emph
  {\bibinfo {booktitle} {Adv. Chem. Phys.}}},\ Vol.\ \bibinfo {volume} {XXVI}\
  (\bibinfo  {publisher} {John Wiley \& Sons, Ltd},\ \bibinfo {year} {1974})\
  pp.\ \bibinfo {pages} {1--106}\BibitemShut {NoStop}%
\bibitem [{\citenamefont {Champagne}, \citenamefont {Li},\ and\ \citenamefont
  {Hunt}(2000)}]{Champagne2k}%
  \BibitemOpen
  \bibfield  {author} {\bibinfo {author} {\bibfnamefont {M.}~\bibnamefont
  {Champagne}}, \bibinfo {author} {\bibfnamefont {X.}~\bibnamefont {Li}}, \
  and\ \bibinfo {author} {\bibfnamefont {K.}~\bibnamefont {Hunt}},\ }\bibfield
  {title} {\enquote {\bibinfo {title} {Nonadditive three-body polarizabilities
  of molecules interacting at long range: Theory and numerical results for the
  inert gases, {H}${}_2$, {N}${}_2$, {CO}${}_2$, and {CH}${}_4$},}\ }\href@noop
  {} {\bibfield  {journal} {\bibinfo  {journal} {J. Chem. Phys.}\ }\textbf
  {\bibinfo {volume} {112}},\ \bibinfo {pages} {1893--1906} (\bibinfo {year}
  {2000})}\BibitemShut {NoStop}%
\bibitem [{\citenamefont {Feynman}\ and\ \citenamefont {Hibbs}(1965)}]{FH}%
  \BibitemOpen
  \bibfield  {author} {\bibinfo {author} {\bibfnamefont {R.~P.}\ \bibnamefont
  {Feynman}}\ and\ \bibinfo {author} {\bibfnamefont {A.}~\bibnamefont
  {Hibbs}},\ }\href@noop {} {\emph {\bibinfo {title} {Quantum Mechanics and
  Path Integrals}}}\ (\bibinfo  {publisher} {McGraw-Hill},\ \bibinfo {address}
  {New York},\ \bibinfo {year} {1965})\BibitemShut {NoStop}%
\bibitem [{\citenamefont {Garberoglio}\ and\ \citenamefont
  {Harvey}(2011)}]{Garberoglio2011a}%
  \BibitemOpen
  \bibfield  {author} {\bibinfo {author} {\bibfnamefont {G.}~\bibnamefont
  {Garberoglio}}\ and\ \bibinfo {author} {\bibfnamefont {A.~H.}\ \bibnamefont
  {Harvey}},\ }\bibfield  {title} {\enquote {\bibinfo {title} {Path-integral
  calculation of the third virial coefficient of quantum gases at low
  temperatures},}\ }\href {\doibase 10.1063/1.3573564} {\bibfield  {journal}
  {\bibinfo  {journal} {J. Chem. Phys.}\ }\textbf {\bibinfo {volume} {134}},\
  \bibinfo {pages} {134106} (\bibinfo {year} {2011})}\BibitemShut {NoStop}%
\bibitem [{\citenamefont {Garberoglio}\ and\ \citenamefont
  {Harvey}(2020{\natexlab{b}})}]{Garberoglio2011aerr}%
  \BibitemOpen
  \bibfield  {author} {\bibinfo {author} {\bibfnamefont {G.}~\bibnamefont
  {Garberoglio}}\ and\ \bibinfo {author} {\bibfnamefont {A.~H.}\ \bibnamefont
  {Harvey}},\ }\bibfield  {title} {\enquote {\bibinfo {title} {Erratum:
  {P}ath-integral calculation of the third virial coefficient of quantum gases
  at low temperatures},}\ }\href@noop {} {\bibfield  {journal} {\bibinfo
  {journal} {J. Chem. Phys.}\ }\textbf {\bibinfo {volume} {152}},\ \bibinfo
  {pages} {199903} (\bibinfo {year} {2020}{\natexlab{b}})}\BibitemShut
  {NoStop}%
\bibitem [{\citenamefont {Garberoglio}(2008)}]{Garberoglio2008}%
  \BibitemOpen
  \bibfield  {author} {\bibinfo {author} {\bibfnamefont {G.}~\bibnamefont
  {Garberoglio}},\ }\bibfield  {title} {\enquote {\bibinfo {title} {Boltzmann
  bias grand canonical {M}onte {C}arlo.}}\ }\href {\doibase 10.1063/1.2883683}
  {\bibfield  {journal} {\bibinfo  {journal} {J. Chem. Phys.}\ }\textbf
  {\bibinfo {volume} {128}},\ \bibinfo {pages} {134109} (\bibinfo {year}
  {2008})}\BibitemShut {NoStop}%
\bibitem [{\citenamefont {Garberoglio}, \citenamefont {Moldover},\ and\
  \citenamefont {Harvey}(2020)}]{Garberoglio2011err}%
  \BibitemOpen
  \bibfield  {author} {\bibinfo {author} {\bibfnamefont {G.}~\bibnamefont
  {Garberoglio}}, \bibinfo {author} {\bibfnamefont {M.~R.}\ \bibnamefont
  {Moldover}}, \ and\ \bibinfo {author} {\bibfnamefont {A.~H.}\ \bibnamefont
  {Harvey}},\ }\bibfield  {title} {\enquote {\bibinfo {title} {Erratum:
  Improved first-principles calculation of the third virial coefficient of
  helium},}\ }\href {\doibase 10.6028/jres.116.016} {\bibfield  {journal}
  {\bibinfo  {journal} {J. Res. Nat. Inst. Stand. Technol.}\ }\textbf {\bibinfo
  {volume} {125}},\ \bibinfo {pages} {125019} (\bibinfo {year}
  {2020})}\BibitemShut {NoStop}%
\bibitem [{\citenamefont {Garberoglio}(2012)}]{Garberoglio2012}%
  \BibitemOpen
  \bibfield  {author} {\bibinfo {author} {\bibfnamefont {G.}~\bibnamefont
  {Garberoglio}},\ }\bibfield  {title} {\enquote {\bibinfo {title} {Quantum
  effects on virial coefficients: a numerical approach using centroids},}\
  }\href {\doibase 10.1016/j.cplett.2012.01.005} {\bibfield  {journal}
  {\bibinfo  {journal} {Chem. Phys. Lett.}\ }\textbf {\bibinfo {volume}
  {525-526}},\ \bibinfo {pages} {19} (\bibinfo {year} {2012})}\BibitemShut
  {NoStop}%
\bibitem [{\citenamefont {Kreckel}(1997)}]{pvegas}%
  \BibitemOpen
  \bibfield  {author} {\bibinfo {author} {\bibfnamefont {R.}~\bibnamefont
  {Kreckel}},\ }\bibfield  {title} {\enquote {\bibinfo {title} {Parallelization
  of adaptive {MC} integrators},}\ }\href {\doibase
  10.1016/S0010-4655(97)00099-4} {\bibfield  {journal} {\bibinfo  {journal}
  {Comp. Phys. Comm.}\ }\textbf {\bibinfo {volume} {106}},\ \bibinfo {pages}
  {258} (\bibinfo {year} {1997})}\BibitemShut {NoStop}%
\bibitem [{\citenamefont {Czachorowski}\ \emph {et~al.}(2020)\citenamefont
  {Czachorowski}, \citenamefont {Przybytek}, \citenamefont {Lesiuk},
  \citenamefont {Puchalski},\ and\ \citenamefont {Jeziorski}}]{u2_2020}%
  \BibitemOpen
  \bibfield  {author} {\bibinfo {author} {\bibfnamefont {P.}~\bibnamefont
  {Czachorowski}}, \bibinfo {author} {\bibfnamefont {M.}~\bibnamefont
  {Przybytek}}, \bibinfo {author} {\bibfnamefont {M.}~\bibnamefont {Lesiuk}},
  \bibinfo {author} {\bibfnamefont {M.}~\bibnamefont {Puchalski}}, \ and\
  \bibinfo {author} {\bibfnamefont {B.}~\bibnamefont {Jeziorski}},\ }\bibfield
  {title} {\enquote {\bibinfo {title} {Second virial coefficients for $^4${He}
  and $^3${He} from an accurate relativistic interaction potential},}\ }\href
  {\doibase 10.1103/PhysRevA.102.042810} {\bibfield  {journal} {\bibinfo
  {journal} {Phys. Rev. A}\ }\textbf {\bibinfo {volume} {102}},\ \bibinfo
  {pages} {042810} (\bibinfo {year} {2020})}\BibitemShut {NoStop}%
\bibitem [{\citenamefont {Cencek}, \citenamefont {Patkowski},\ and\
  \citenamefont {Szalewicz}(2009)}]{FCI}%
  \BibitemOpen
  \bibfield  {author} {\bibinfo {author} {\bibfnamefont {W.}~\bibnamefont
  {Cencek}}, \bibinfo {author} {\bibfnamefont {K.}~\bibnamefont {Patkowski}}, \
  and\ \bibinfo {author} {\bibfnamefont {K.}~\bibnamefont {Szalewicz}},\
  }\bibfield  {title} {\enquote {\bibinfo {title}
  {Full-configuration-interaction calculation of three-body nonadditive
  contribution to helium interaction potential},}\ }\href@noop {} {\bibfield
  {journal} {\bibinfo  {journal} {J. Chem. Phys.}\ }\textbf {\bibinfo {volume}
  {131}},\ \bibinfo {pages} {064105} (\bibinfo {year} {2009})}\BibitemShut
  {NoStop}%
\bibitem [{\citenamefont {Cencek}, \citenamefont {Komasa},\ and\ \citenamefont
  {Szalewicz}(2011)}]{Cencek11}%
  \BibitemOpen
  \bibfield  {author} {\bibinfo {author} {\bibfnamefont {W.}~\bibnamefont
  {Cencek}}, \bibinfo {author} {\bibfnamefont {J.}~\bibnamefont {Komasa}}, \
  and\ \bibinfo {author} {\bibfnamefont {K.}~\bibnamefont {Szalewicz}},\
  }\bibfield  {title} {\enquote {\bibinfo {title} {Collision-induced dipole
  polarizability of helium dimer from explicitly correlated calculations},}\
  }\href {\doibase 10.1063/1.3603968} {\bibfield  {journal} {\bibinfo
  {journal} {J. Chem. Phys.}\ }\textbf {\bibinfo {volume} {135}},\ \bibinfo
  {pages} {014301} (\bibinfo {year} {2011})}\BibitemShut {NoStop}%
\bibitem [{\citenamefont {Huot}\ and\ \citenamefont {Bose}(1991)}]{Huot91}%
  \BibitemOpen
  \bibfield  {author} {\bibinfo {author} {\bibfnamefont {J.}~\bibnamefont
  {Huot}}\ and\ \bibinfo {author} {\bibfnamefont {T.~K.}\ \bibnamefont
  {Bose}},\ }\bibfield  {title} {\enquote {\bibinfo {title} {Experimental
  determination of the dielectric virial coefficients of atomic gases as a
  function of temperature},}\ }\href {\doibase 10.1063/1.461801} {\bibfield
  {journal} {\bibinfo  {journal} {J. Chem. Phys.}\ }\textbf {\bibinfo {volume}
  {95}},\ \bibinfo {pages} {2683--2687} (\bibinfo {year} {1991})}\BibitemShut
  {NoStop}%
\bibitem [{\citenamefont {Kirouac}\ and\ \citenamefont
  {Bose}(1976)}]{Kirouac76}%
  \BibitemOpen
  \bibfield  {author} {\bibinfo {author} {\bibfnamefont {S.}~\bibnamefont
  {Kirouac}}\ and\ \bibinfo {author} {\bibfnamefont {T.~K.}\ \bibnamefont
  {Bose}},\ }\bibfield  {title} {\enquote {\bibinfo {title} {Polarizability and
  dielectric properties of helium},}\ }\href {\doibase 10.1063/1.432383}
  {\bibfield  {journal} {\bibinfo  {journal} {J. Chem. Phys.}\ }\textbf
  {\bibinfo {volume} {64}},\ \bibinfo {pages} {1580--1582} (\bibinfo {year}
  {1976})}\BibitemShut {NoStop}%
\bibitem [{\citenamefont {White}\ and\ \citenamefont
  {Gugan}(1992)}]{White_1992}%
  \BibitemOpen
  \bibfield  {author} {\bibinfo {author} {\bibfnamefont {M.~P.}\ \bibnamefont
  {White}}\ and\ \bibinfo {author} {\bibfnamefont {D.}~\bibnamefont {Gugan}},\
  }\bibfield  {title} {\enquote {\bibinfo {title} {Direct measurements of the
  dielectric virial coefficients of ${}^4${H}e between 3 {K} and 18 {K}},}\
  }\href {\doibase 10.1088/0026-1394/29/1/006} {\bibfield  {journal} {\bibinfo
  {journal} {Metrologia}\ }\textbf {\bibinfo {volume} {29}},\ \bibinfo {pages}
  {37--57} (\bibinfo {year} {1992})}\BibitemShut {NoStop}%
\bibitem [{\citenamefont {Hellmann}\ \emph {et~al.}(2021)\citenamefont
  {Hellmann}, \citenamefont {Gaiser}, \citenamefont {Fellmuth}, \citenamefont
  {Vasyltsova},\ and\ \citenamefont {Bich}}]{Hellmann21}%
  \BibitemOpen
  \bibfield  {author} {\bibinfo {author} {\bibfnamefont {R.}~\bibnamefont
  {Hellmann}}, \bibinfo {author} {\bibfnamefont {C.}~\bibnamefont {Gaiser}},
  \bibinfo {author} {\bibfnamefont {B.}~\bibnamefont {Fellmuth}}, \bibinfo
  {author} {\bibfnamefont {T.}~\bibnamefont {Vasyltsova}}, \ and\ \bibinfo
  {author} {\bibfnamefont {E.}~\bibnamefont {Bich}},\ }\bibfield  {title}
  {\enquote {\bibinfo {title} {Thermophysical properties of low-density neon
  gas from highly accurate first-principles calculations and
  dielectric-constant gas thermometry measurements},}\ }\href@noop {}
  {\bibfield  {journal} {\bibinfo  {journal} {J. Chem. Phys.}\ }\textbf
  {\bibinfo {volume} {154}},\ \bibinfo {pages} {164304} (\bibinfo {year}
  {2021})}\BibitemShut {NoStop}%
\bibitem [{\citenamefont {Schwerdtfeger}\ and\ \citenamefont
  {Hermann}(2009)}]{Schwerdtfeger09}%
  \BibitemOpen
  \bibfield  {author} {\bibinfo {author} {\bibfnamefont {P.}~\bibnamefont
  {Schwerdtfeger}}\ and\ \bibinfo {author} {\bibfnamefont {A.}~\bibnamefont
  {Hermann}},\ }\bibfield  {title} {\enquote {\bibinfo {title} {Equation of
  state for solid neon from quantum theory},}\ }\href {\doibase
  10.1103/PhysRevB.80.064106} {\bibfield  {journal} {\bibinfo  {journal} {Phys.
  Rev. B}\ }\textbf {\bibinfo {volume} {80}},\ \bibinfo {pages} {064106}
  (\bibinfo {year} {2009})}\BibitemShut {NoStop}%
\bibitem [{\citenamefont {{Hellmann
  (Helmut-Schmidt-Universit{\"a}t/Universit{\"a}t der Bundeswehr
  Hamburg)}}(2021)}]{Hellmann_pc}%
  \BibitemOpen
  \bibfield  {author} {\bibinfo {author} {\bibfnamefont {R.}~\bibnamefont
  {{Hellmann (Helmut-Schmidt-Universit{\"a}t/Universit{\"a}t der Bundeswehr
  Hamburg)}}},\ }\href@noop {} {} (\bibinfo {year} {2021}),\ \bibinfo {note}
  {private communication}\BibitemShut {NoStop}%
\bibitem [{\citenamefont {Patkowski}\ and\ \citenamefont
  {Szalewicz}(2010)}]{Patkowski10}%
  \BibitemOpen
  \bibfield  {author} {\bibinfo {author} {\bibfnamefont {K.}~\bibnamefont
  {Patkowski}}\ and\ \bibinfo {author} {\bibfnamefont {K.}~\bibnamefont
  {Szalewicz}},\ }\bibfield  {title} {\enquote {\bibinfo {title} {Argon pair
  potential at basis set and excitation limits},}\ }\href {\doibase
  10.1063/1.3478513} {\bibfield  {journal} {\bibinfo  {journal} {J. Chem.
  Phys.}\ }\textbf {\bibinfo {volume} {133}},\ \bibinfo {pages} {094304}
  (\bibinfo {year} {2010})}\BibitemShut {NoStop}%
\bibitem [{\citenamefont {Vogel}\ \emph {et~al.}(2010)\citenamefont {Vogel},
  \citenamefont {J{\"a}ger}, \citenamefont {Hellmann},\ and\ \citenamefont
  {Bich}}]{Vogel10}%
  \BibitemOpen
  \bibfield  {author} {\bibinfo {author} {\bibfnamefont {E.}~\bibnamefont
  {Vogel}}, \bibinfo {author} {\bibfnamefont {B.}~\bibnamefont {J{\"a}ger}},
  \bibinfo {author} {\bibfnamefont {R.}~\bibnamefont {Hellmann}}, \ and\
  \bibinfo {author} {\bibfnamefont {E.}~\bibnamefont {Bich}},\ }\bibfield
  {title} {\enquote {\bibinfo {title} {{\em Ab initio} pair potential energy
  curve for the argon atom pair and thermophysical properties for the dilute
  argon gas. {II}. {T}hermophysical properties for low-density argon},}\ }\href
  {\doibase 10.1080/00268976.2010.507557} {\bibfield  {journal} {\bibinfo
  {journal} {Mol. Phys.}\ }\textbf {\bibinfo {volume} {108}},\ \bibinfo {pages}
  {3335--3352} (\bibinfo {year} {2010})}\BibitemShut {NoStop}%
\bibitem [{\citenamefont {Cencek}\ \emph {et~al.}(2013)\citenamefont {Cencek},
  \citenamefont {Garberoglio}, \citenamefont {Harvey}, \citenamefont
  {McLinden},\ and\ \citenamefont {Szalewicz}}]{Cencek2013}%
  \BibitemOpen
  \bibfield  {author} {\bibinfo {author} {\bibfnamefont {W.}~\bibnamefont
  {Cencek}}, \bibinfo {author} {\bibfnamefont {G.}~\bibnamefont {Garberoglio}},
  \bibinfo {author} {\bibfnamefont {A.~H.}\ \bibnamefont {Harvey}}, \bibinfo
  {author} {\bibfnamefont {M.~O.}\ \bibnamefont {McLinden}}, \ and\ \bibinfo
  {author} {\bibfnamefont {K.}~\bibnamefont {Szalewicz}},\ }\bibfield  {title}
  {\enquote {\bibinfo {title} {Three-body nonadditive potential for argon with
  estimated uncertainties and third virial coefficient},}\ }\href {\doibase
  10.1021/jp4018579} {\bibfield  {journal} {\bibinfo  {journal} {J. Phys. Chem.
  A}\ }\textbf {\bibinfo {volume} {117}},\ \bibinfo {pages} {7542--7552}
  (\bibinfo {year} {2013})}\BibitemShut {NoStop}%
\bibitem [{\citenamefont {Achtermann}, \citenamefont {Magnus},\ and\
  \citenamefont {Bose}(1991)}]{Achtermann91}%
  \BibitemOpen
  \bibfield  {author} {\bibinfo {author} {\bibfnamefont {H.~J.}\ \bibnamefont
  {Achtermann}}, \bibinfo {author} {\bibfnamefont {G.}~\bibnamefont {Magnus}},
  \ and\ \bibinfo {author} {\bibfnamefont {T.~K.}\ \bibnamefont {Bose}},\
  }\bibfield  {title} {\enquote {\bibinfo {title} {Refractivity virial
  coefficients of gaseous $\mathrm{CH}_4$, $\mathrm{C}_2\mathrm{H}_4$,
  $\mathrm{C}_2\mathrm{H}_6$, $\mathrm{CO}_2$, $\mathrm{SF}_6$, $\mathrm{H}_2$,
  $\mathrm{N}_2$, {H}e, and {A}r},}\ }\href {\doibase 10.1063/1.460478}
  {\bibfield  {journal} {\bibinfo  {journal} {J. Chem. Phys.}\ }\textbf
  {\bibinfo {volume} {94}},\ \bibinfo {pages} {5669--5684} (\bibinfo {year}
  {1991})}\BibitemShut {NoStop}%
\bibitem [{\citenamefont {Achtermann}\ \emph {et~al.}(1993)\citenamefont
  {Achtermann}, \citenamefont {Hong}, \citenamefont {Magnus}, \citenamefont
  {Aziz},\ and\ \citenamefont {Slaman}}]{Achtermann93}%
  \BibitemOpen
  \bibfield  {author} {\bibinfo {author} {\bibfnamefont {H.~J.}\ \bibnamefont
  {Achtermann}}, \bibinfo {author} {\bibfnamefont {J.~G.}\ \bibnamefont
  {Hong}}, \bibinfo {author} {\bibfnamefont {G.}~\bibnamefont {Magnus}},
  \bibinfo {author} {\bibfnamefont {R.~A.}\ \bibnamefont {Aziz}}, \ and\
  \bibinfo {author} {\bibfnamefont {M.~J.}\ \bibnamefont {Slaman}},\ }\bibfield
   {title} {\enquote {\bibinfo {title} {Experimental determination of the
  refractivity virial coefficients of atomic gases},}\ }\href {\doibase
  10.1063/1.464212} {\bibfield  {journal} {\bibinfo  {journal} {J. Chem.
  Phys.}\ }\textbf {\bibinfo {volume} {98}},\ \bibinfo {pages} {2308--2318}
  (\bibinfo {year} {1993})}\BibitemShut {NoStop}%
\bibitem [{\citenamefont {Bose}\ \emph {et~al.}(1988)\citenamefont {Bose},
  \citenamefont {Boudjarane}, \citenamefont {Huot},\ and\ \citenamefont
  {St-Arnaud}}]{Bose88}%
  \BibitemOpen
  \bibfield  {author} {\bibinfo {author} {\bibfnamefont {T.~K.}\ \bibnamefont
  {Bose}}, \bibinfo {author} {\bibfnamefont {K.}~\bibnamefont {Boudjarane}},
  \bibinfo {author} {\bibfnamefont {J.}~\bibnamefont {Huot}}, \ and\ \bibinfo
  {author} {\bibfnamefont {J.~M.}\ \bibnamefont {St-Arnaud}},\ }\bibfield
  {title} {\enquote {\bibinfo {title} {Refractivity virial coefficients of
  $\mathrm{C}_2\mathrm{H}_4$ and $\mathrm{C}_2\mathrm{H}_4$--{A}r mixtures},}\
  }\href {\doibase 10.1063/1.455273} {\bibfield  {journal} {\bibinfo  {journal}
  {J. Chem. Phys.}\ }\textbf {\bibinfo {volume} {89}},\ \bibinfo {pages}
  {7435--7440} (\bibinfo {year} {1988})}\BibitemShut {NoStop}%
\bibitem [{\citenamefont {Moldover}\ and\ \citenamefont
  {Buckley}(2001)}]{Moldover2001}%
  \BibitemOpen
  \bibfield  {author} {\bibinfo {author} {\bibfnamefont {M.~R.}\ \bibnamefont
  {Moldover}}\ and\ \bibinfo {author} {\bibfnamefont {T.~J.}\ \bibnamefont
  {Buckley}},\ }\bibfield  {title} {\enquote {\bibinfo {title} {Reference
  values of the dielectric constant of natural gas components determined with a
  cross capacitor},}\ }\href {\doibase 10.1023/A:1010731117103} {\bibfield
  {journal} {\bibinfo  {journal} {Int. J. Thermophys.}\ }\textbf {\bibinfo
  {volume} {22}},\ \bibinfo {pages} {859--885} (\bibinfo {year}
  {2001})}\BibitemShut {NoStop}%
\end{thebibliography}%

\end{document}